\documentclass[12pt,a4paper,final]{iopart}

\usepackage{iopams}  
\usepackage{graphicx}

\pdfoutput=1
\expandafter\let\csname equation*\endcsname\relax
\expandafter\let\csname endequation*\endcsname\relax
\usepackage{amsfonts,amssymb,amsmath}
\usepackage[usenames,dvipsnames]{xcolor}
\usepackage{xspace}
\usepackage{tikz}
\usetikzlibrary{shapes,arrows}
\usetikzlibrary{matrix}
\usetikzlibrary{shapes.multipart}
\usetikzlibrary{er,positioning}
\usepackage[normalem]{ulem} 
\usepackage{url}
\usepackage{empheq}
\usepackage{float}
\usepackage{cancel}
\usepackage{xcolor}
\usepackage{soul}
\usepackage[caption=false]{subfig}
\usepackage[T1]{fontenc}
\usepackage[utf8]{inputenc}
\usepackage{microtype}

\newcommand{\eps}{\ensuremath{\varepsilon}}

\newcommand{\pd}{\partial}

\newcommand{\nn}{\nonumber}

\newcommand{\txt}[1]{{\textrm{\tiny{#1}}}}
\newcommand{\mpl}{\ensuremath{m_\txt{pl}}}

\newcommand{\dual}{\,{}^*\!}         

\begin{document}

\title[Numerical black hole initial data and shadows in dynamical Chern-Simons gravity]{Numerical black hole initial data and shadows in dynamical Chern-Simons gravity}

\author{Maria Okounkova$^{1}$}

    \address{$^1$Theoretical Astrophysics,
    Walter Burke Institute for Theoretical Physics,\\
    California Institute of Technology, Pasadena, CA 91125, USA}
\ead{mokounko@tapir.caltech.edu}

\author{Mark A. Scheel$^{1}$}

\author{Saul A. Teukolsky$^{1,2}$}

\address{$^2$Center for Astrophysics and
    Planetary Science, Cornell University, Ithaca, New York 14853, USA}

\date{\today}

\begin{abstract}
We present a scheme for generating first-order metric perturbation initial data for an arbitrary background and source. We then apply this scheme to derive metric perturbations in order-reduced dynamical Chern-Simons gravity (dCS). In particular, we solve for metric perturbations on a black hole background that are sourced by a first-order dCS scalar field. This gives us the leading-order metric perturbation to the spacetime in dCS gravity. We then use these solutions to compute black hole shadows in the linearly perturbed spacetime by evolving null geodesics. We present a novel scheme to decompose the shape of the shadow into multipoles parametrized by the spin of the background black hole and the perturbation parameter $\eps^2$. We find that we can differentiate the presence of a pure Kerr spacetime from a spacetime with a dCS perturbation using the shadow, allowing in part for a null-hypothesis test of general relativity. We then consider these results in the context of the Event Horizon Telescope. 
\end{abstract}

\section*{Response to Referee 1}


\section{Introduction} 

Einstein's theory of general relativity (GR) has passed all precision tests to date~\cite{Will:2014kxa}. In particular, model-independent 
tests using binary black hole merger data from the Laser Interferometry Gravitational Wave Observatory (LIGO) are consistent with GR at the 96\% confidence level~\cite{Abbott:2017vtc, TheLIGOScientific:2016src, Yunes:2016jcc}. 

However, at some length scale GR must be reconciled with quantum mechanics in a theory of quantum 
gravity. Black holes and black hole binaries probe the strong-field, non-linear, high-curvature regime of gravity, and thus observations 
of these systems might contain signatures of quantum gravity. Our goal is to predict these signatures. 

We know from the first LIGO detections that deviations from GR are small, and thus rather than considering black holes in a 
fully quantum theory, we can calculate their properties in \textit{effective field theories} (EFTs). These theories involve adding perturbative quantum-gravity-motivated terms to the 
Einstein-Hilbert action of general relativity. Since these theories are classical, we can hope to apply
the numerical tools used to study GR (a classical theory) to these quantum-gravity-motivated theories. 

One such EFT is dynamical Chern-Simons gravity (dCS), which modifies the action of GR through the inclusion of a scalar field
coupled to spacetime curvature~\cite{Alexander:2009tp}. In particular, this theory has motivations in string theory~\cite{Green:1984sg}, 
loop quantum gravity~\cite{Taveras:2008yf,Mercuri:2009zt}, and inflation~\cite{Weinberg:2008hq}. The full effective field theory, however, most likely does not have a well-posed initial value formulation~\cite{Delsate:2014hba}. However, we can expand the theory around general relativity in order to guarantee a well-posed system of equations at each order~\cite{Okounkova:2017yby}. This is in part justified by the first LIGO detection, which found deviations from GR in black hole systems to be small~\cite{TheLIGOScientific:2016src}.  In a previous study, we investigated the leading-order behavior of the dCS scalar field in a binary black hole system,
quantifying the amount by which gravitational waves in dCS gravity would differ from those in pure GR~\cite{Okounkova:2017yby}. 

In this study, we numerically compute \textit{metric perturbations} in dCS. In other words, we calculate to leading order the modifications to 
a pure GR spacetime due to the presence of the dCS scalar field. Such modifications will be required , for example, as initial data
to perform binary black hole simulations involving a dCS metric perturbation.  We thus produce and test a formalism for generating metric perturbation initial 
data based on the extended conformal thin sandwich formalism (cf.~\cite{baumgarteShapiroBook} for a review). Previous studies have
considered such modifications, but we present the first such formalism that can be used in the binary black hole case~\cite{Yunes:2009hc, Yagi:2012ya, 
McNees:2015srl, Delsate:2018ome, Cardenas-Avendano:2018ocb, Ayzenberg:2018jip}. 

In addition to LIGO, an instrument coming online that will have the power to probe the strong-field regime of gravity is the Event Horizon Telescope (EHT). The primary goal of this instrument (a very long baseline interferometry array of radio telescopes) is to image black hole event horizons, including those of Sgr A*, the black hole at the center of the Milky Way galaxy, and the black hole of the center of the M87 galaxy~\cite{Ricarte:2014nca,Falcke:2018uqa}. The EHT in part has the power to image the \textit{black hole shadow}, a dark region on the image corresponding to angles at which no photons reach the observer, because of light-bending and the presence of an event horizon. The shadow, for a black hole with a given mass and spin, has a precise shape predicted by GR, and thus deviations from this shape can be used to test the theory~\cite{Mizuno:2018lxz,Psaltis:2018xkc,Mizuno:2018lxz}. Since the paths of photons are determined by the spacetime itself, resolving the shadow corresponds to directly probing the metric of the spacetime,
and hence is a metric test of GR. Moreover, predictions for black hole shadows exist in other theories of gravity. Thus one can go beyond performing a null-hypothesis test of GR and instead test specific theories. Additionally, since the mass of Sgr A* is ${}\sim10^6\,M_\odot$, whereas the masses of black holes observed by LIGO are ${}\sim10\,M_\odot$, the EHT probes gravity on a wholly new scale~\cite{Baker:2014zba}. 

Given dCS metric perturbations, our goal is  to compute the black hole shadow in a dCS-modified spacetime, and quantify the effects (including degeneracies) on the shape of the shadow as a function of mass, spin, and the dCS coupling parameter. We can then estimate whether the EHT would be able to 
resolve these deviations.

\subsection{Roadmap and conventions}
This paper is organized as follows. In Sec.~\ref{sec:Solving}, we derive and provide all of the equations for the formalism for generating metric perturbation initial data. In Sec.~\ref{sec:dCSData}, we specifically apply this formalism to black holes in dCS gravity, presenting convergent initial data results. In Sec.~\ref{sec:dCSPhysics}, we present results using stationary dCS metric perturbation initial data to calculate black hole shadows. We conclude in Sec.~\ref{sec:Conclusion}. 

We set $G = c = 1$ throughout. Quantities are given in terms of  units of $M$, the ADM mass of the system. Latin letters in the beginning of the alphabet $\{a, b, c, d \ldots \}$ denote 4-dimensional spacetime indices, while Latin letters in the middle of the alphabet $\{i, j, k, l \ldots \}$ denote 3-dimensional spatial indices. $\psi_{ab}$ refers to the spacetime metric, while $g_{ij}$ refers to the spatial metric from a 3+1 decomposition with corresponding timelike unit normal one-form $n_a$ (cf.~\cite{baumgarteShapiroBook} for a review of the 3+1 ADM formalism).

\section{Solving for general metric perturbation initial data} 
\label{sec:Solving}

\subsection{Overview}
\label{sec:SolvingOverview}

In standard numerical general relativity, initial data is often generated using the extended conformal thin sandwich formalism~\cite{Cook2004,  Pfeiffer:2005zm, Lovelace2008, Lovelace2009, Ossokine:2015yla}. A thorough review of this method is presented in~\cite{baumgarteShapiroBook} and a derivation is presented in~\cite{Pfeiffer:2004nc}. This formalism decomposes the 3+1 ADM Hamiltonian and momentum constraints, as well as the equation for the time derivative of the extrinsic curvature, to generate a set of elliptic equations to numerically solve for initial data. 

Recall that in the 3+1 decomposition, the constraints and time derivative of the extrinsic curvature are given as
\begin{align}
\label{eq:Ham}
&R + K^2 - K_{ij} K^{ij} = 16 \pi \rho \,, \\
\label{eq:Mom}
&D_j (K^{ij} - g^{ij} K) = 8 \pi S^i \,, \\
\label{eq:dtK} 
&\pd_t K_{ij} = \alpha (R_{ij} - 2 K_{ij} K^k {}_j + K K_{ij}) - D_i D_j \alpha \\
\nn & \quad  \quad \quad  - 8 \pi \alpha (S_{ij} - \frac{1}{2} g_{ij} (S - \rho)) + \beta^k \pd_k K_{ij} + K_{ik} \pd_j \beta^k + K_{kj} \pd_i \beta^k, 
\end{align}
where $g_{ij}$ is the spatial metric with corresponding covariant derivative $D_i$, $\alpha$ is the lapse, and $\beta^i$ is the shift. 
$K_{ij}$ is the extrinsic curvature with trace $K$, and $R_{ij}$ is the spatial Ricci tensor with trace $R$. 
The matter terms $\rho$, $S^i$, $S_{ij}$, and $S$ are defined with respect to the 
stress-energy tensor $T_{ab}$ and timelike unit normal one-form $n_a$ as 
\begin{align}
\label{eq:rho}
\rho &\equiv n_a n_b T^{ab}\,, \\
\label{eq:Si}
S^i &\equiv -g^{ij} n^a T_{aj} \,, \\
\label{eq:Sij}
S_{ij} &\equiv g_{ia} g_{jb} T^{ab} \,, \\
\label{eq:S}
S &\equiv g^{ij} S_{ij} \,,
\end{align}
where the time-space components of the spatial metric are given via $g_{ab} \equiv \psi_{ab} + n_a n_b$ for spacetime metric $\psi_{ab}$. 

The extended conformal thin sandwich formalism involves writing the spatial metric  in terms of a conformal metric $\bar{g}_{ij}$ as
\begin{align}
\label{eq:gConformal}
g_{ij} = \psi^4 \bar{g}_{ij}\,,
\end{align}
where $\psi$ is known as the conformal factor. Additionally, the time derivative of the spatial metric is decomposed as
\begin{align}
\label{eq:dtgConformal}
 u_{ij} &=  \pd_t g_{ij} - \frac{2}{3} g_{ij} (-\alpha K + D_i \beta^i)\,,
\end{align}
where the function $u_{ij}$ is related to the time derivative of the conformal metric as
\begin{align}
\label{eq:uConformal}
u_{ij} = \psi^4 \bar{u}_{ij}\,,
\end{align}
with
\begin{align}
\label{eq:Udef}
 \bar{u}_{ij} \equiv \pd_t \bar{g}_{ij} \,.
\end{align}

In this formalism, the extrinsic curvature is decomposed into traceless and trace parts as
\begin{align}
\label{eq:KDecomposed}
K_{ij} = A_{ij} + \frac{1}{3} g_{ij} K \,,
\end{align}
where $A_{ij}$ is the traceless part of $K_{ij}$, and is conformally transformed as
\begin{align}
\label{eq:AConformal}
A_{ij} = \psi^{-2} \bar{A}_{ij} \,,
\end{align}
with
\begin{align}
\label{eq:Aij}
\bar{A}^{ij} &= \frac{\psi^7}{2\alpha \psi} ((\bar{L}\beta)^{ij} - \bar{u}^{ij}) \,, \\
\label{eq:LShift}
(\bar{L} \beta)^{ij} &\equiv \bar{D}^i \beta^j + \bar{D}^j \beta^i - \frac{2}{3} \bar{g}^{ij} \bar{D}_k \beta^k \,.
\end{align}
Here, $\bar{D}_i$ refers to the covariant derivative with respect to the conformal metric, $\bar{g}_{ij}$. 

Having defined all of these quantities, we can now recast Eqs.~\eqref{eq:Ham},~\eqref{eq:Mom} and~\eqref{eq:dtK} to give an elliptic equation for the conformal factor, 
\begin{align}
\label{eq:PsiXCTS}
\bar{D}^2 \psi - \frac{1}{8} \psi \bar{R} - \frac{1}{12} \psi^5 K^2 + \frac{1}{8} \psi^{-7} \bar{A}_{ij} \bar{A}^{ij} = -2 \pi \psi^5 \rho\,,
\end{align}
an elliptic equation for the shift,
\begin{align}
\label{eq:ShiftXCTS}
(\bar{\Lambda}_L \beta)^i - (\bar{L} \beta)^{ij} \bar{D}_j \ln \bar{\alpha} &= \bar{\alpha} \bar{D}_j (\bar{\alpha}^{-1} \bar{u}^{ij}) + \frac{4}{3} \bar{\alpha} \psi^6 \bar{D}^i K + 16 \pi \bar{\alpha} \psi^{10} S^i \,,
\end{align}
and an elliptic equation for $\alpha \psi$,
\begin{align}
\label{eq:LapseXCTS}
\bar{D}^2 (\alpha \psi) &= \alpha \psi(\frac{7}{8} \psi^{-8} \bar{A}_{ij} \bar{A}^{ij} + \frac{5}{12} \psi^4 K^2 + \frac{1}{8} \bar{R} \\
\nn & \quad + 2\pi \psi^4 (\rho + 2 S)) - \psi^5 \pd_t K + \psi^5 \beta^i \bar{D}_i K\,.
\end{align}
Here, $\bar{\alpha} \equiv \psi^{-6} \alpha$ is the densitized lapse, $\bar{R}$ is the conformal Ricci scalar computed for $\bar{g}_{ij}$, and $(\bar{\Lambda}_L\beta)^i$ is the vector Laplacian (cf.~\cite{baumgarteShapiroBook}). 

In the extended conformal thin sandwich formalism, we are freely allowed to specify
\begin{align}
\label{eq:FreeData}
\boxed{\textrm{Free data: } \bar{g}_{ij}, \bar{u}_{ij}, K, \pd_t K} \,,
\end{align}
and solve for the variables 
\begin{align}
\label{eq:SolvedData}
\boxed{\textrm{Solved data: }  \psi, \beta^i, \alpha\psi} \,.
\end{align}

We are interested in solving for initial data for linear metric perturbations of the form
\begin{align}
\label{eq:MetricPert}
\psi_{ab} \to \psi_{ab} + \Delta \psi_{ab}\,.
\end{align}
In order to solve for perturbed initial data, we will perturb the extended conformal thin sandwich equations. Our overall goal is to perturb each of these equations to linear order, which will give us elliptic equations for the perturbed variables with the same principal part as the background equations. Throughout, we will denote by $\Delta X$ the first-order (linear) perturbation to some variable $X$.  We perturb each of the variables as 
\begin{align}
\label{eq:ConformalPeturbed}
\psi &\to \psi + \Delta \psi \,, \\
\beta^i &\to \beta^i +  \Delta \beta^i \,, \\
\alpha \psi &\to \alpha \psi + (\alpha \Delta \psi + \Delta \alpha \psi) \,,
\end{align}
and solve for $\Delta \psi$, the perturbation to the conformal factor, $\Delta \beta^i$, the perturbation to the shift, and
\begin{align}
\Delta C \equiv \Delta(\alpha \psi) = \alpha \Delta \psi + \Delta \alpha \psi\,,
\end{align}
the perturbation to the lapse times the conformal factor. 

The equations will additionally involve perturbing metric quantities to first order, such as
\begin{align}
\bar{g}_{ij} &\to \bar{g}_{ij} +  \Delta \bar{g}_{ij}\,, \\
\bar{u}_{ij} &\to \bar{u}_{ij} +  \Delta \bar{u}_{ij}\,, \\
K &\to K + \Delta K\,,\\
\pd_t K &\to \pd_t K + \pd_t \Delta K\,,
\end{align}
where $\Delta \bar{u}_{ij} \equiv \pd_t \Delta \bar{g}_{ij}$.
We outline these terms in more detail in~\ref{sec:PerturbedQuantitiesAppendix}.

Much like we have the solved data and free data in the extended conformal thin sandwich formalism, we will have
\begin{align}
\boxed{\textrm{Perturbed free data: } \Delta \bar{g}_{ij}, \Delta \bar{u}_{ij}, \Delta K, \pd_t \Delta K}\,,
\end{align}
and
\begin{align}
\boxed{\textrm{Perturbed solved data: }  \Delta \psi, \Delta \beta^i, \Delta C}\,.
\end{align}

\subsection{Perturbed initial data formalism} 
\label{sec:PerturbedFormalism}

We now perturb Eqs.~\eqref{eq:PsiXCTS},~\eqref{eq:ShiftXCTS}, and~\eqref{eq:LapseXCTS} to obtain elliptic equations for $\Delta \psi$, $\Delta \beta^i$, and $\Delta C$. Each of these equations involves the perturbations to the extended conformal thin sandwich quantities. For example, the equations will include the first-order perturbation to $\bar{A}^{ij}$ (defined in Eq.~\eqref{eq:Aij}), denoted $\Delta \bar{A}^{ij}$. We leave the derivations of the perturbations to all of the extended conformal thin sandwich quantities to~\ref{sec:PerturbedQuantitiesAppendix}, and present the perturbations to the elliptic equations for $\Delta \psi$, $\Delta \beta^i$, and $\Delta C$ here.  

\subsubsection{Perturbed equations} 
\label{sec:PerturbedEquations}

Perturbing Eq.~\eqref{eq:PsiXCTS} yields an elliptic equation for
$\Delta \psi$.
We obtain 
\begin{align}
\label{eq:DeltaPsiEq}
0 &= -\bar{D}^2 \Delta \psi  - \Delta ( \bar{D}^2) \psi \\
\nn & \quad + \frac{1}{8} \Delta \psi \bar{R} + \frac{1}{8} \psi \Delta \bar{R}  + \frac{5}{12} \psi^4 \Delta \psi K^2 + \frac{1}{6} \psi^5 K \Delta K \\
\nn & \quad +\frac{7}{8} \psi^{-8} \Delta \psi \bar{A}_{ij} \bar{A}^{ij} - \frac{1}{8} \psi^{-7} (\Delta \bar{A}_{ij} \bar{A}^{ij} +  \bar{A}_{ij} \Delta \bar{A}^{ij}) \\
\nn & \quad - 2\pi(5 \psi^4 \Delta \psi \rho + \psi^5 \Delta \rho)\,,
\end{align}
where $\bar{D}^2 \Delta \psi$ is the principal part of this perturbed equation. 

Perturbing Eq.~\eqref{eq:LapseXCTS} yields an elliptic equation for $\Delta C$. Since this equation is longer, we will do it piece by piece, 
splitting the original expression as 

\begin{align}
0 &= \underbrace{- \bar{D}^2 (\alpha \psi)}_\textrm{Principal part}  + \underbrace{\alpha \psi \left (\frac{7}{8} \psi^{-8} \bar{A}_{ij} \bar{A}^{ij} + \frac{5}{12} \psi^4 K^2 + \frac{1}{8} \bar{R}\right)}_\textrm{Non-matter terms} \\
\nn & \quad \underbrace{- \psi^5 \pd_t K + \psi^5 \beta^i \bar{D}_i K}_\textrm{Non-matter terms} + \underbrace{\alpha \psi 2\pi \psi^4 (\rho + 2 S)}_\textrm{Matter terms} \,.
\end{align}
Perturbing the \textbf{Matter terms}, we obtain
\begin{align}
\label{eq:Cmatter}
\Delta (\textrm{$C$ Matter terms}) = 2\pi(\Delta C \psi^4 (\rho + 2 S) + 4 \alpha \psi  \psi^3 \Delta \psi (\rho + 2 S) + \alpha \psi  \psi^4 (\Delta \rho + 2 \Delta S)) \,.
\end{align}
Next, perturbing the \textbf{Non-matter terms}, we obtain
\begin{align}
\label{eq:CNonmatter}
\Delta (\textrm{$C$ Non-matter terms}) = & \Delta C \left (\frac{7}{8} \psi^{-8} \bar{A}_{ij} \bar{A}^{ij} + \frac{5}{12} \psi^4 K^2 + \frac{1}{8} \bar{R}\right) \\
\nn & + \alpha \psi (-7 \psi^{-9} \Delta \psi \bar{A}_{ij} \bar{A}^{ij} +  \frac{7}{8} \psi^{-8} (\Delta \bar{A}_{ij} \bar{A}^{ij} + \bar{A}_{ij} \Delta \bar{A}^{ij})\\
\nn & + \frac{5}{3} \psi^3 \Delta \psi K^2 + \frac{5}{6} \psi^4 K \Delta K +  \frac{1}{8} \Delta \bar{R} ) \\
\nn & - 5 \psi^4 \Delta \psi \pd_t K - \psi^5 \pd_t \Delta K \\
\nn & + 5 \psi^4 \Delta \psi \beta^i \bar{D}_i K +  \psi^5 \Delta \beta^i \bar{D}_i K  +  \psi^5 \beta^i \bar{D}_i \Delta K \,.
\end{align}
Finally, for the perturbation to the \textbf{Principal part}, we obtain
\begin{align}
\label{eq:CPP}
\Delta (\textrm{$C$ Principal part}) = - \bar{D}^2 (\Delta C) - \Delta (\bar{D}^2) (\alpha \psi) \,,
\end{align}
where the first term gives us the principal part for the perturbed equation. We combine these terms into an overall elliptic equation for $\Delta C$
\begin{align}
\label{eq:DeltaCEq}
\Delta (\textrm{$C$ Principal part}) + \Delta (\textrm{$C$ Non-matter terms}) + \Delta (\textrm{$C$ Matter terms}) = 0\,,
\end{align}
where the perturbed terms are given in Eqs.~\eqref{eq:CPP},~\eqref{eq:Cmatter}, and~\eqref{eq:CNonmatter}.

In order to complete our system of equations, we perturb Eq.~\eqref{eq:ShiftXCTS} to obtain an equation for $\Delta \beta^i$. 
In practice, we solve the momentum constraint with the principal part
\begin{align}
-\alpha \psi \bar{D}_j \left( \frac{1}{\alpha \psi} (\bar{L}\beta)^{ij} \right) \,,
\end{align}
where the momentum constraint has been rewritten using as

\begin{align}
\label{eq:TransformedShiftXCTS}
0 &= - \alpha\psi \bar{D}_j \left( \frac{1}{\alpha\psi} (\bar{L}\beta)^{ij} \right) \\
\nn & + \bar{D}_j \bar{u}^{ij} -  \frac{14 \alpha \psi} {\psi^8} \bar{A}^{ij} \bar{D}_j \psi -  \bar{u}^{ij}  \frac{\bar{D}_j \alpha \psi}{\alpha \psi}  + \frac{4}{3} \frac{\alpha \psi}{\psi} \bar{D}^i K \\
\nn & + 16 \pi \alpha \psi \psi^{3} S^i \,.
\end{align}

For simplicity, we split up Eq.~\eqref{eq:TransformedShiftXCTS} as 
\begin{align}
0 &= \underbrace{- \alpha\psi \bar{D}_j \left( \frac{1}{\alpha\psi} (\bar{L}\beta)^{ij} \right)}_\textrm{Principal part} \\
\nn & \underbrace{+ \bar{D}_j \bar{u}^{ij} -  \frac{14 \alpha \psi} {\psi^8} \bar{A}^{ij} \bar{D}_j \psi -  \bar{u}^{ij}  \frac{\bar{D}_j \alpha \psi}{\alpha \psi}  + \frac{4}{3} \frac{\alpha \psi}{\psi} \bar{D}^i K}_\textrm{Non-matter terms} \\
\nn & \underbrace{+ 16 \pi \alpha \psi \psi^{3} S^i}_\textrm{Matter terms} \,. 
\end{align}
Perturbing the \textbf{Matter terms}, we obtain
\begin{align}
\label{eq:BMatter}
\Delta(\textrm{$\beta^i$ Matter terms}) = 16 \pi (\Delta C \psi^3 S^i + 3  \alpha \psi \psi^2 \Delta \psi S^i +  \alpha \psi \psi^{3} \Delta S^i) \,.
\end{align}
Perturbing the \textbf{Non-matter terms} gives
\begin{align}
\label{eq:BNonmatter}
\Delta(\textrm{$\beta^i$ Non-matter terms}) =  &\Delta (\bar{D})_j \bar{u}^{ij} + \bar{D}_j \Delta \bar{u}^{ij}  \\
\nn & -  \frac{14 \Delta C} {\psi^8} \bar{A}^{ij} \bar{D}_j \psi +  \frac{112 \alpha \psi} {\psi^9} \Delta \psi \bar{A}^{ij} \bar{D}_j \psi  \\
\nn &-  \frac{14 \alpha \psi} {\psi^8}(\Delta \bar{A}^{ij} \bar{D}_j \psi  + \bar{A}^{ij}  \bar{D}_j \Delta \psi) \\ 
\nn & - \Delta \bar{u}^{ij}  \frac{\bar{D}_j \alpha \psi}{\alpha \psi}  - \frac{\bar{u}^{ij}}{\alpha \psi} \bar{D}_j \Delta C + \bar{u}^{ij}  \Delta C \frac{\bar{D}_j \alpha \psi}{(\alpha \psi)^2}\\
\nn & + \frac{4}{3} \frac{\Delta C}{\psi} \bar{D}^i K - \frac{4}{3} \frac{\alpha \psi}{\psi^2} \Delta \psi \bar{D}^i K\\
\nn & +  \frac{4}{3} \frac{\alpha \psi}{\psi} ( \Delta (\bar{D})^j K + \bar{D}^i \Delta K) \,.
\end{align}
Finally, perturbing the \textbf{Principal part} gives 
\begin{align}
\label{eq:BPP}
\Delta(\textrm{$\beta^i$ Principal part}) = &- \alpha\psi \bar{D}_j \left( \frac{1}{\alpha\psi} ((\bar{L}\Delta \beta)^{ij} + (\Delta(\bar{L})\beta)^{ij} )  \right) \\
\nn &- \Delta (\bar{D})_j (\bar{L}\beta)^{ij}  + \frac{ (\bar{L}\beta)^{ij}}{\alpha \psi} \bar{D}_j  \Delta C -  (\bar{L}\beta)^{ij}  \frac{\Delta C}{(\alpha \psi)^2} \bar{D}_j  \alpha \psi \,.
\end{align}
Our overall elliptic equation for $\Delta \beta^i$ is 
\begin{align}
\label{eq:DeltaShiftEq}
\Delta (\textrm{$\beta^i$ Principal part}) + \Delta (\textrm{$\beta^i$ Non-matter terms}) + \Delta (\textrm{$\beta^i$ Matter terms}) = 0 \,,
\end{align}
where the perturbed terms are given in Eqs.~\eqref{eq:BPP},~\eqref{eq:BMatter}, and~\eqref{eq:BNonmatter}.

Thus, we have derived a set of three second-order, elliptic equations for $\Delta \psi$, $\Delta C$, and $\Delta \beta^i$. We solve Eq.~\eqref{eq:DeltaPsiEq} for $\Delta \psi$, Eq.~\eqref{eq:DeltaCEq} for $\Delta C$, and Eq.\eqref{eq:DeltaShiftEq} for $\Delta \beta^i$. The principal parts of all of these equations are the same as in the unperturbed extended conformal thin sandwich equations. Thus, for numerical solutions, we can reuse the preconditioning matrices and linearized operators that are used in the unperturbed equations. The specific details of the numerical computation can be found in~\cite{Pfeiffer:2005zm}. 

\subsubsection{Reconstructing perturbed data}
\label{sec:DataReconstruction}

Given solutions of the equations from the previous section for $\Delta \psi, \Delta C, \Delta \beta^i$, as well as the perturbed free data and background data, we now wish to reconstruct $\Delta g_{ij}$, the full perturbed spatial metric, and $\pd_t \Delta g_{ij}$, its time derivative. This allows us to construct $\Delta \psi_{ab}$, the perturbation to the spacetime metric, and its time derivative, $\pd_t \Delta \psi_{ab}$. We detail this procedure in~\ref{sec:ReconstructionAppendix}.

\subsubsection{Constraint satisfaction}
\label{sec:Constraints}

Writing down the perturbed initial data equations is only the first half of the problem. In practice, we need to make sure that solving them produces data that satisfies the Hamiltonian and momentum constraints. In the unperturbed case, we simply check that Eqs.~\eqref{eq:Ham} and~\eqref{eq:Mom} are satisfied. In the perturbed case, since we are computing a \textit{linear perturbation}, we do not expect the full, non-linear constraints to be satisfied. Rather, the first-order linearization of these constraints should hold. We thus perturb these constraints to give 
\begin{align}
\label{eq:PertHam}
\Delta H &\equiv \Delta R + 2 K \Delta K  - \Delta K_{ij} K^{ij} - K_{ij} \Delta  K^{ij} - 16 \pi \Delta \rho\,,
\end{align}
for the perturbed Hamiltonian constraint, and 
\begin{align}
\label{eq:PertMom}
\Delta M_i &\equiv \Delta g^{jk}(D_j K_{ki} - D_i K_{jk}) \\
\nn &\quad +  g^{jk}(\Delta (D)_j K_{ki} - \Delta (D)_i K_{jk} + D_j \Delta K_{ki} - D_i \Delta K_{jk}) - 8 \pi \Delta S_i
\end{align}
for the perturbed momentum constraint. 
	Constraint-satisfying perturbed initial data will thus have $\Delta H = 0$ and $\Delta M_i = 0$.
	
	In practice, these conditions will never be exactly satisfied, but  we can check that these quantities tend toward zero with increasing numerical resolution. In our case, we use a spectral code~\cite{SpECwebsite}, and thus the constraint violation converges to zero exponentially. In order to give meaning to the level of constraint violation, we normalize each constraint by the magnitude of the fields contained therein. 

\subsection{Boundary conditions}
\label{sec:PerturbedBCs}

 Before solving elliptic equations for metric perturbations for a generic source $\Delta T_{ab}$, we must impose boundary conditions. Specifically, we must impose conditions on $\Delta \psi$, $\Delta C$, and $\Delta \beta^i$ at spatial infinity $(R \to \infty)$. In our spectral code~\cite{SpECwebsite}, we excise the black hole singularities from the computational domain via a surface that conforms to the apparent horizon (or is slightly inside the apparent horizon)~\cite{Hemberger:2012jz}. Thus, for a background containing a black hole, we must specify boundary conditions on the excision surface. In the case of a black hole binary, there are two such excision surfaces, one for each hole, and thus we must specify boundary conditions on each of them. 

Let us now consider the boundary conditions we would impose in the case where the background spacetime contains a single black hole. First, the matter distribution, and hence the source of the perturbation, should decay at least as fast as $1/R^2$ as $R \to \infty$. Thus, we can choose the conditions
\begin{align}
\Delta \psi |_{r \to \infty} &= 0,\\
 \Delta \beta^i |_{r \to \infty} &= 0, \\
\Delta  C |_{r \to \infty} &= 0\,.
\end{align}
These conditions agree with  the perturbed boundary conditions for an isolated black hole spacetime given in~\cite{Cook2004,Pfeiffer:2005zm}. In practice, we extend the (finite) outer domain to $R = 10^{14}\,M$, more than sufficient to satisfy these conditions.

For conditions on the inner boundaries, which correspond to apparent horizons, we impose the set of apparent horizon boundary conditions for $\psi$, $\alpha$, and $\beta^i$ given in~\cite{Cook2004,Pfeiffer:2005zm}. The conditions ensure that the surface has zero expansion, and has a desired value for the spin. In our case, we can perturb these apparent horizon boundary conditions to give conditions on $\Delta \psi$, $\Delta C$, and $\Delta \beta^i$. 

Specifically, for the unperturbed boundary conditions, the condition on $\psi$ corresponds to setting the expansion of the surface to be zero, the condition on $\beta^i$ corresponds to setting the spin and also setting the shear of the null rays on the horizon to be zero, while the condition on $\alpha$ is physically unconstrained and can be set with a Dirichlet condition. The condition on $\psi$ takes the form 

\begin{align}
\label{eq:PsiBC}
0 &= -\bar{P}^i \pd_i \psi - B \psi + \frac{1}{8}  \frac{\psi^4}{\alpha \psi}  (C_{ij}) (\bar{L}\beta^{ij} - u^{ij}) + \frac{\psi^3}{12} C^{ij}\bar{g}_{ij} K \,, 
\end{align}
where 

\begin{align}
N &\equiv \sqrt{\bar{g}^{ij} \hat{n}_i \hat{n}_j} \,, \\
\bar{P}^i &\equiv \frac{\hat{n}_j \bar{g}^{ij}}{N} \,,
\end{align}
with $\hat{n}^i$ being the normal vector to the inner boundary, and

\begin{align}
C^{ij} &\equiv \bar{g}^{ij} - \bar{P}^i \bar{P}^j \,, \\
B &\equiv \frac{1}{4 N}(C^{ij}) (\pd_j \hat{n}_i - \bar{\Gamma}^l_{ij} \hat{n}_l) \,.
\end{align}

When perturbing this condition, we must consider what to do with the perturbation to $\hat{n}^i$. If we set $\Delta \hat{n}^i = 0$, then the excision surface corresponds to a horizon for the background, and the overall shape of the surface is not perturbed. By choosing a non-zero $\Delta \hat{n}^i$, we can, for example, set the expansion of the background metric plus the first-order metric perturbation to zero, and hence have the surface correspond to a linearly perturbed horizon. In this study, we set $\Delta \hat{n}^i = 0$ for simplicity. 

Perturbing Eq.~\eqref{eq:PsiBC}, we thus obtain
\begin{align}
\label{eq:DeltaPsiBC}
0 &= - \Delta \bar{P}^i \pd_i \psi - \bar{P}^i \pd_i \Delta \psi - \Delta B \psi - B \Delta \psi \\
\nn & \quad  + \frac{1}{2} \frac{\psi^3 \Delta \psi}{\alpha \psi} (C_{ij}) (\bar{L}\beta^{ij} - u^{ij}) \\
\nn & \quad - \frac{1}{8} \frac{\psi^4}{(\alpha \psi)^2} \Delta C (C_{ij}) (\bar{L}\beta^{ij} - u^{ij}) \\
\nn & \quad  + \frac{1}{8} \frac{\psi^4}{\alpha \psi} (\Delta C_{ij}) (\bar{L}\beta^{ij} - u^{ij}) \\
\nn & \quad  + \frac{1}{8} \frac{\psi^4}{\alpha \psi} (C_{ij}) (\Delta (\bar{L}\beta^{ij}) - \Delta u^{ij}) \\
\nn & \quad  + \frac{\psi^2 \Delta \psi }{4} C^{ij}\bar{g}_{ij} K  +\frac{\psi^3}{12} \Delta C^{ij}\bar{g}_{ij} K +\frac{\psi^3}{12} C^{ij} \Delta \bar{g}_{ij} K  \\
\nn & \quad  +\frac{\psi^3}{12} C^{ij}\bar{g}_{ij} \Delta K
\end{align}
on the excision surface, where
\begin{align}
\Delta N &=  \frac{1}{2N} \Delta \bar{g}^{ij} \hat{n}_i \hat{n}_j  \,,  \\
\Delta \bar{P}^i &= \frac{\hat{n}_j \Delta \bar{g}^{ij}}{N} - \frac{\hat{n}_j \bar{g}^{ij}}{N^2} \Delta N \,, \\
\Delta C^{ij} &= \Delta \bar{g}^{ij} - \Delta \bar{P}^i \bar{P}^j - \bar{P}^i \Delta \bar{P}^j \,,  \\
\Delta B &= -\frac{1}{4 N^2} \Delta N (C^{ij}) (\pd_j \hat{n}_i - \bar{\Gamma}^l_{ij} \hat{n}_l) \,, \\
\nn & \quad + \frac{1}{4 N}(\Delta C^{ij}) (\pd_j \hat{n}_i - \bar{\Gamma}^l_{ij} \hat{n}_l)  \\
\nn & \quad + \frac{1}{4 N}(C^{ij}) (- \Delta \bar{\Gamma}^l_{ij} \hat{n}_l) \,.
\end{align}

Next, the background boundary condition on $\beta^i$ takes the form 
\begin{align}
\label{eq:ShiftBC}
0 &= \beta^i - \frac{1}{\psi^3}  \frac{\hat{n}_j g^{ij}}{N} \alpha \psi  - \xi^i
\end{align}
on the inner boundary. Here, $\xi^i$ is the vector 
\begin{align}
 \xi^i =  \Omega_x X^i + \Omega_y Y^i +  \Omega_z Z^i \,,
\end{align}
where $\Omega_i$ corresponds to the components of the orbital angular momentum, and $X^i$, $Y^i$, and $Z^i$ have the form 
\begin{align}
X^i &= (0, -z, y)  \,, \\
Y^i &= (z, 0, -x) \,, \\
Z^i &= (-y, x, 0)\,.
\end{align}
Now, when we perturb this condition, we must consider how to perturb $\Omega_i$. Setting this to a non-zero value gives a spin to the metric perturbation as well. 

Perturbing Eq.~\eqref{eq:ShiftBC}, we thus obtain
\begin{align}
\label{eq:DeltaShiftBC}
0 &= \Delta \beta^i + 3\frac{1}{\psi^4} \Delta \psi \frac{\hat{n}_j g^{ij}}{N} \alpha \psi  \\
\nn & -  \frac{1}{\psi^3} \frac{\hat{n}_j \Delta g^{ij}}{N} \alpha \psi \\
\nn & +  \frac{1}{\psi^3} \frac{\hat{n}_j g^{ij}}{N^2} \Delta N \alpha \psi \\
\nn & -  \frac{1}{\psi^3} \frac{\hat{n}_j g^{ij}}{N} \Delta C \\
\nn & - \Delta \xi^i
\end{align}
on the excision surface, where $\Delta \xi^i$  is the vector
\begin{align}
\Delta \xi^i = \Delta \Omega_x X^i + \Delta \Omega_y Y^i + \Delta \Omega_z Z^i \,.
\end{align}

The Dirichlet boundary condition on $\alpha$, meanwhile, can be perturbed to give a Dirichlet boundary condition on $\Delta C$. However, we are already solving Eq.~\eqref{eq:DeltaPsiBC} for $\Delta \psi$, and thus to uncouple these equations, we can instead try to drive $\Delta \alpha$ to some desired value $\Delta \alpha_\mathrm{Desired}$ on the excision surface via the Dirichlet condition
\begin{align}
\label{eq:DeltaCBC}
0 = \Delta C - (\Delta \psi \alpha + \psi \Delta \alpha_\mathrm{Desired})\,.
\end{align}

We can generalize the isolated black hole case to a binary black hole case, by applying Eqs.~\eqref{eq:DeltaPsiBC}~\eqref{eq:DeltaShiftBC}, and~\eqref{eq:DeltaCBC} to each excision surface corresponding to a horizon, and applying a boost in the case of an initial velocity. 

\subsection{Summary}

Thus, in order to generate metric perturbation initial data given some source $\Delta T_{ab}$ and background spacetime metric $\psi_{ab}$, we solve the elliptic equations given in Sec.~\ref{sec:PerturbedEquations} for $\Delta \psi$, $\Delta C$, and $\Delta \beta^i$. We then apply the formulae in Sec.~\ref{sec:DataReconstruction} to construct $\Delta \psi_{ab}$, the perturbed spacetime metric for these variables. For the case where the background is an isolated black hole, we can apply the perturbed version of the horizon boundary conditions on $\Delta \psi$, $\Delta C$, and $\Delta \beta^i$ given in Sec.~\ref{sec:PerturbedBCs}. In order to generate stationary data on an isolated black hole background, we choose $\Delta \Omega_i$ in Eq.~\eqref{eq:DeltaShiftBC} to be equal to the $\Omega_i$ of the background.

Note that, as outlined in Sec.~\ref{sec:SolvingOverview}, we have the freedom to choose $\Delta \bar{g}_{ij}$, $\Delta \bar{u}_{ij}$, $\Delta K$, and $\pd_t \Delta K$. To simplify the calculation in the isolated black hole case, we choose $\Delta \bar{g}_{ij} = 0$, and thus $\Delta g_{ij} = 4 \psi^3 \Delta \psi \bar{g}_{ij}$. For stationarity, we choose $\Delta \bar{u}_{ij} = 0$ and $\pd_t \Delta K = 0$ to set as many time derivatives to zero as possible. We similarly choose $\Delta K = 0$. 

\section{Solving for metric perturbations in dCS} 
\label{sec:dCSData}

\subsection{Order reduction scheme}
\label{sec:OrderReduction}

We now turn to applying the method for solving for metric perturbation initial data outlined in Sec.~\ref{sec:Solving} to isolated black holes in dynamical Chern-Simons (dCS) gravity. The dCS action for a metric $\psi_{ab}$ and scalar field $\vartheta$ is given by
\begin{align}
\label{eq:dCSAction}
\int d^4 x \sqrt{-\psi} \left( \frac{\mpl^2}{2} R - \frac{1}{2} (\pd \vartheta)^2 - \frac{\mpl}{8} \ell^2 \vartheta \dual RR \right) \,,
\end{align}
where $\ell$ is a coupling constant with dimensions of length,
\begin{align}
\dual RR \equiv \dual R^{abcd} R_{abcd}
\end{align}
is the Pontryagin density, where $\dual R^{abcd} = \frac{1}{2} \epsilon^{abef} R_{ef}{}^{cd}$ is the dual of the Riemann tensor and $\epsilon^{abcd} \equiv -[abcd]/\sqrt{-\psi}$ is the fully-antisymmetric Levi-Civita tensor, and $\mpl$ is the Planck mass. 

Varying the action in Eq.~\eqref{eq:dCSAction}, we obtain a sourced wave equation for the scalar field,
\begin{align}
\square \vartheta = \frac{\mpl \ell^2}{8} \dual RR \,,
\end{align}
where $\square \equiv \nabla_a \nabla^a$ is the d'Alembertian operator. For the metric, we obtain a corrected Einstein field equation
\begin{align}
\mpl^2 G_{ab} + \mpl \ell^2 C_{ab} = T_{ab} \,,
\end{align}
where $T_{ab}$ is the kinetic stress-energy tensor of $\vartheta$,
\begin{align}
\label{eq:KineticT}
T_{ab} = \nabla_a \vartheta \nabla_b \vartheta - \frac{1}{2} \psi_{ab} \nabla_c \vartheta \nabla^c \vartheta \,,
\end{align}
and 
\begin{align}
\label{eq:CTensor}
C_{ab} \equiv \epsilon_{cde(a} \nabla^d R_{b)}{}^c \nabla^e \vartheta + \dual  R^c{}_{(ab)}{}^d \nabla_c \nabla_d \vartheta\,.
\end{align}

Note that $C_{ab}$ contains third derivatives of the metric, and thus these equations of motion must likely not have a well-posed  initial value problem~\cite{Delsate:2014hba}. However, in the perturbation limit we can solve these equations of motion using an \textit{order reduction scheme}, expanding the metric and scalar field in powers of a parameter $\varepsilon$ that counts powers of $\ell^2$:
\begin{align}
\psi_{ab} &= \psi_{ab}^{(0)} + \sum_{k = 1}^\infty \eps^k h_{ab}^{(k)} \,, \\
\vartheta &= \sum_{k = 0}^\infty \eps^k \vartheta^{(k)} \,.
\end{align}
The key is that at each order of this scheme, we will obtain equations of motion with the same principal part as GR. Perturbing around GR is justified in part by the first LIGO detection, which showed that deviations from GR in black hole systems are small~\cite{TheLIGOScientific:2016src}. 

At zeroth order in $\eps$, we obtain for our equations of motion
\begin{align}
\mpl^2 G_{ab} [\psi^{(0)}] &= T_{ab}^{(0)} \,, \\
\square^{(0)} \vartheta^{(0)} = 0 \,,
\end{align}
where $T_{ab}^{(0)}$ is the stress-energy tensor constructed from $\vartheta^{(0)}$. 
Since the zeroth order scalar field has no source, we can take $\vartheta^{(0)} = 0$.
This is turn means that the equation for the metric at zeroth order is a pure GR Einstein field equation. 

 At first order, meanwhile, we obtain the equation
\begin{align}
\label{eq:FirstOrderField}
\square^{(0)} \vartheta^{(1)} = \frac{\mpl}{8} \ell^2 [\dual RR]^{(0)}
\end{align}
for the first-order scalar field $\vartheta^{(1)}$, and the equation
\begin{align}
\label{eq:FirstOrderMetric}
\mpl^2 G_{ab}[h^{(1)}] = -\mpl \ell^2 C_{ab}^{(0)} + T_{ab}^{(1)}
\end{align}
for the first-order metric perturbation, where $G_{ab}$ is the Einstein-Hilbert operator of the background acting on the metric perturbation. Here, $C_{ab}^{(0)}$ is the background value of the tensor defined in Eq.~\eqref{eq:CTensor}, and $T_{ab}^{(1)}$ is the first-order perturbation to the stress-energy tensor given in Eq.~\eqref{eq:KineticT}. However, both $C_{ab}^{(0)}$ and $T_{ab}^{(1)}$ are linear in $\vartheta^{(0)}$, which vanishes, and hence $-\mpl \ell^2 C_{ab}^{(0)} + T_{ab}^{(1)}$, the RHS  of 
Eq.~\eqref{eq:FirstOrderMetric} vanishes, leaving an unsourced metric perturbation,
\begin{align}
\mpl^2 G_{ab}[h^{(1)}] = 0 \,.
\end{align}
Thus, at first order in $\eps$, $h^{(1)}=0$, there is no modification to the metric, and the scalar field is governed by Eq.~\eqref{eq:FirstOrderField}. Indeed, in~\cite{Okounkova:2017yby}, we evolved this $\eps^1$ system on a binary black hole background.

We now turn to order $\varepsilon^2$, where we obtain a metric perturbation sourced by $\vartheta^{(1)}$. Specifically, we obtain
\begin{align}
\mpl^2 G_{ab} [h^{(2)}] = - \mpl \ell^2 C_{ab}^{(1)} [\vartheta^{(1)}] + T_{ab}^{(2)} [\vartheta^{(1)}, \vartheta^{(1)}] \,.
\end{align}
Here, the first term on the right-hand side is  the perturbed $C$-tensor formed from the background metric and the non-vanishing first-order scalar field $\vartheta^{(1)}$ (and hence is non-zero).  The second term is the second-order perturbation to the stress-energy tensor, quadratic in $\vartheta^{(1)}$, and hence also non-zero. 

To simplify the equations and to more easily use the results of the previous section, it is useful to define a new variable $\Psi$ by 
\begin{align}
\vartheta^{(1)} \equiv \frac{\mpl}{8} \ell^2 \Psi\,,
\end{align}
which gives, at first-order
\begin{align}
\label{eq:KGPsiEq}
\square \Psi = \dual RR\,.
\end{align}
Here all metric variables now correspond to the background (in other words, $\dual R R = [\dual R R]^{(0)}$, for example). Similarly, let $\Delta\psi_{ab}$ correspond to the second-order metric perturbation by defining
\begin{align}
h_{ab}^{(2)} \equiv \frac{\ell^4}{8} \Delta \psi_{ab}\,.
\end{align}
The equation for the metric perturbation is thus 
\begin{align}
G_{ab}[\Delta \psi_{ab}] = T^{\mathrm{eff}}_{ab}(\Psi)\,,
\end{align}
where
\begin{align}
\label{eq:Teff}
T_{ab}^{\mathrm{eff}}(\Psi) \equiv -C_{ab} (\Psi) + \frac{1}{8}T_{ab} (\Psi)\,.
\end{align}
We can then write the  $C$-tensor and matter terms in the form
\begin{align}
C_{ab}(\Psi) &= \epsilon_{cde(a} \nabla^d R_{b)}{}^c \nabla^e \Psi + \dual R^c{}_{(ab)}{}^d \nabla_c \nabla_d \Psi \,, \\
T_{ab}(\Psi) &= \nabla_a \Psi \nabla_b \Psi - \frac{1}{2} \psi_{ab} \nabla_c \Psi \nabla^c \Psi \,.
\end{align}
The first term of $C_{ab}$ vanishes when working on a vacuum GR background. 

Thus, $\Delta \psi_{ab}$ is governed by the Einstein tensor and is a perturbation off a GR background of the form $\psi_{ab} \to \psi_{ab} + \Delta \psi_{ab}$ with source $T^{\mathrm{eff}}_{ab}$. Comparing this to Eq.~\eqref{eq:MetricPert}, we can thus use the formalism developed in Sec.~\ref{sec:Solving} to solve for $\Delta \psi_{ab}$ sourced by $T^{\mathrm{eff}}_{ab}$ on a black hole background. 

\subsection{Scalar field initial data}

Before solving for $\Delta \psi_{ab}$, however, we need a scalar field $\Psi$ on a black hole background that obeys Eq.~\eqref{eq:KGPsiEq}. Moreover, in order to obtain stationary data for $\Delta \psi_{ab}$, we require that $\Psi$ is stationary. Rather than attempting to find an analytical solution, we use the numerical solution for $\Psi$ computed using the methods in~\cite{Stein:2014xba}. This solution is valid for any spin. However, this solution is expressed in Boyer-Lindquist coordinates, which are singular at the horizon, and thus we transform to Kerr-Schild coordinates.  The transformation to Kerr-Schild coordinates is given, e.g., in~\cite{MTW}.

We check that the solution for $\Psi$ is constraint satisfying, and moreover that it is stationary. Note that the solution given in~\cite{Stein:2014xba} has its own inherent resolution in terms of the number of radial and angular basis functions. Including more radial basis functions in this solution increases its stationarity. We interpolate the solution onto our grid, generally with a different resolution.

Given this solution for $\Psi$, we then construct the perturbed source terms of Eqs.~\eqref{eq:DeltaRho}~\eqref{eq:DeltaSi}~\eqref{eq:DeltaSij} and~\eqref{eq:DeltaS} using $\Delta T_{ab} = T_{ab}^{\mathrm{eff}}$ computed from $\Psi$ via Eq.~\eqref{eq:Teff}.

\subsection{dCS metric perturbation results} 
\label{sec:MetricResults}

\begin{figure}
  \includegraphics[width=0.8\textwidth]{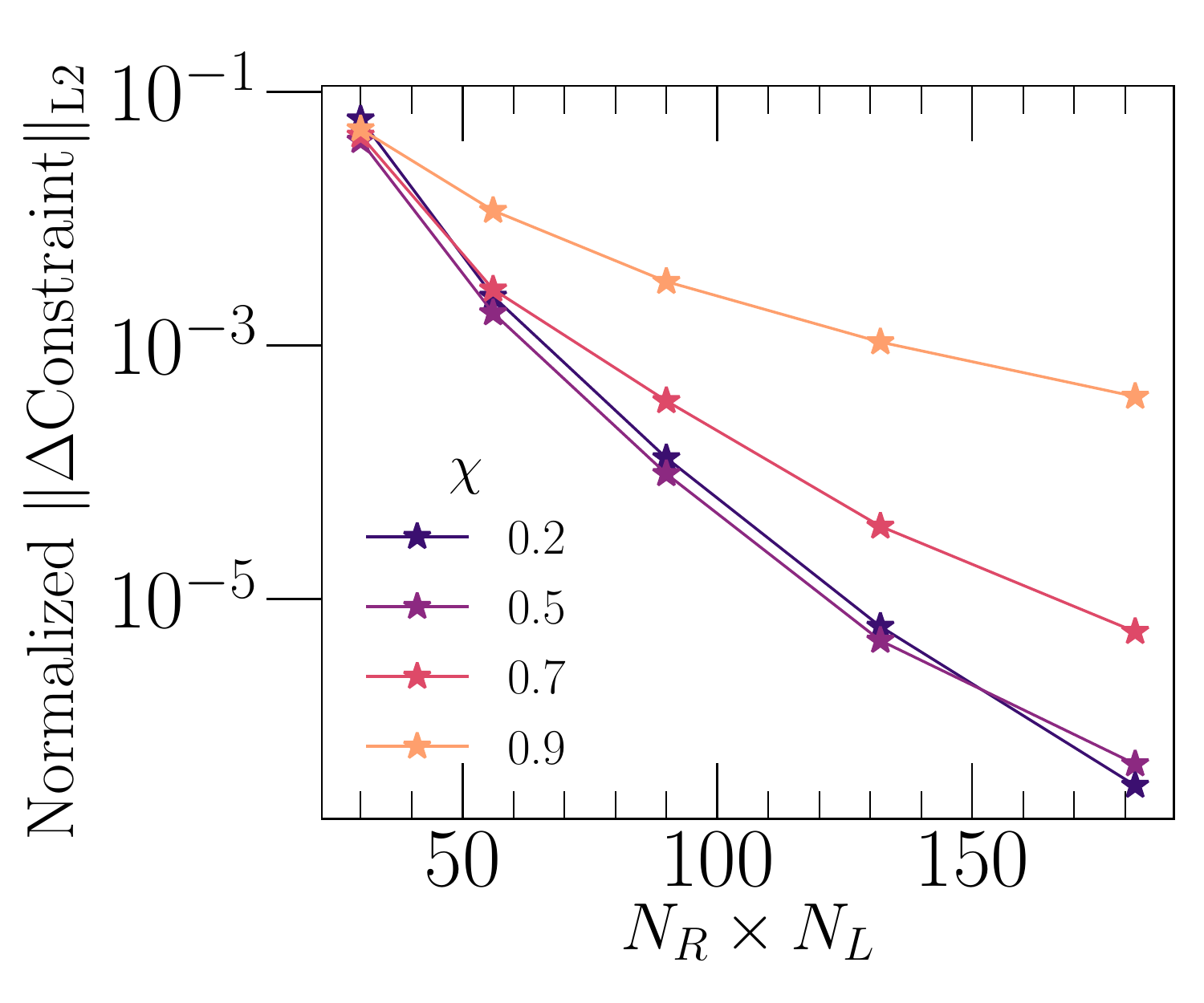}
  \caption{
Convergence of the perturbed constraints  with resolution for a metric perturbation $\Delta \psi_{ab}$ on a Kerr background with given dimensionless spin $\chi$. We evaluate the constraints on the entire numerical grid. The horizontal axis is the number of radial basis functions $N_R$ times angular basis functions $N_L$ in a representative subdomain of our numerical grid. As this number increases, the constraint violation exponentially converges to zero. Higher-spin black holes require more grid points to achieve the same level of constraint satisfaction in the metric perturbation as lower-spin black holes, just as for the unperturbed background spacetime. 
  \label{fig:EllipticSolvingConvergence}}
\end{figure}

Given these source terms, we then apply the formalism developed in Sec.~\ref{sec:Solving} to solve for $\Delta \psi_{ab}$. We verify that our results are convergent by checking the perturbed constraints given in Sec.~\ref{sec:Constraints}.  We solve for the data on a set of nested spherical shells extending from the apparent horizon to $R = 10^{14}\,M$, all with equal numbers of spectral collocation points. Fig.~\ref{fig:EllipticSolvingConvergence} presents the behavior of the normalized, perturbed Hamiltonian and momentum constraints with increasing resolution. The figure shows the exponential convergence of the constraints to zero as the numerical resolution increases.  Higher spins in Kerr-Schild coordinates require more grid points to fully resolve the solution, and thus have a slower level of convergence. Recall likewise that we wish to solve for stationary initial data. In practice, the stationarity converges with increasing resolution. However, at the same numerical resolution, a lower spin will have a greater stationarity, as measured by $\|\Delta g_{ij}\| / \|g_{ij}\|$, than a higher spin. Thus, when comparing quantities across spins in practice, we choose resolutions that give the same level of non-stationarity to mitigate these spin-dependent effects.

\begin{figure}
\centering
\includegraphics[width=0.8\textwidth]{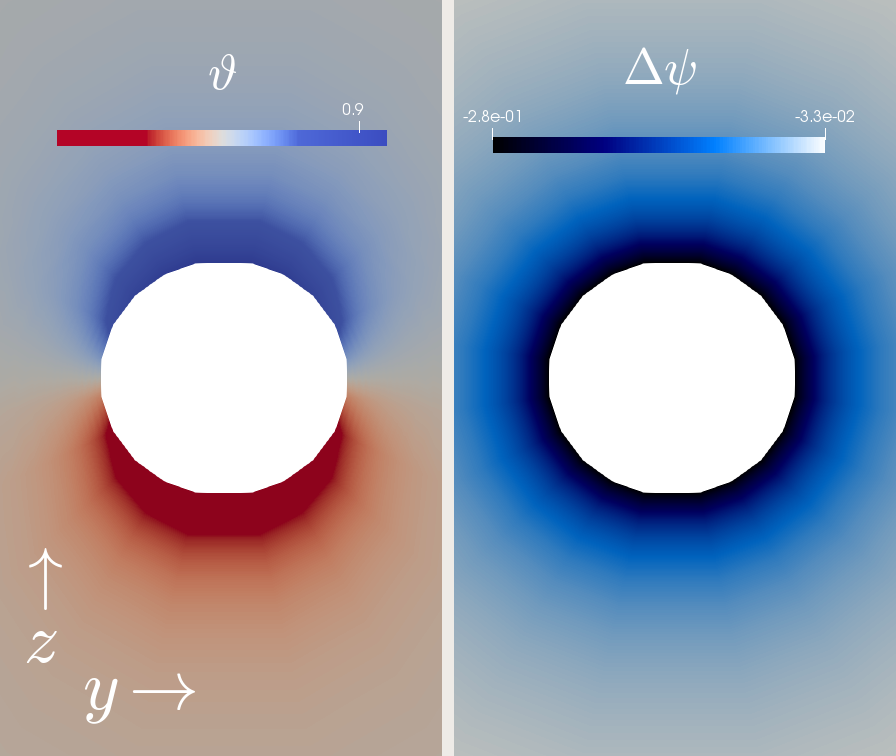}
\caption{Plot of the numerical solution for $\Psi$ from~\cite{Stein:2014xba} (left) and perturbed conformal factor $\Delta \psi$ (right) on a spin $\chi = 0.6$ black hole background, shown in the $y$-$z$ plane. Note that the solution is axisymmetric about the $z$-axis.}
\label{fig:Fields}
\end{figure}

In summary, we have constraint-satisfying data for the second-order metric perturbation in order-reduced dCS gravity. In Fig.~\ref{fig:Fields}, we plot the profiles for the scalar field $\Psi$ as well as the conformal factor $\Delta \psi$. 

The extended conformal thin sandwich formalism can potentially suffer from ill-posedness and non-uniqueness problems if the equations do not have a positive-definite linearization~\cite{Baumgarte:2006ug, PhysRevD.91.042003}. In our case, however, we do not see the appearance of non-unique solutions.  

\section{Physics with dCS metric perturbations} 
\label{sec:dCSPhysics}

We now consider what physics we can extract from these solutions for $\Delta \psi_{ab}$ in dCS. 

\subsection{Regime of validity}
\label{sec:ROV}

\begin{figure}
  \centering
  \includegraphics[width=0.8\textwidth]{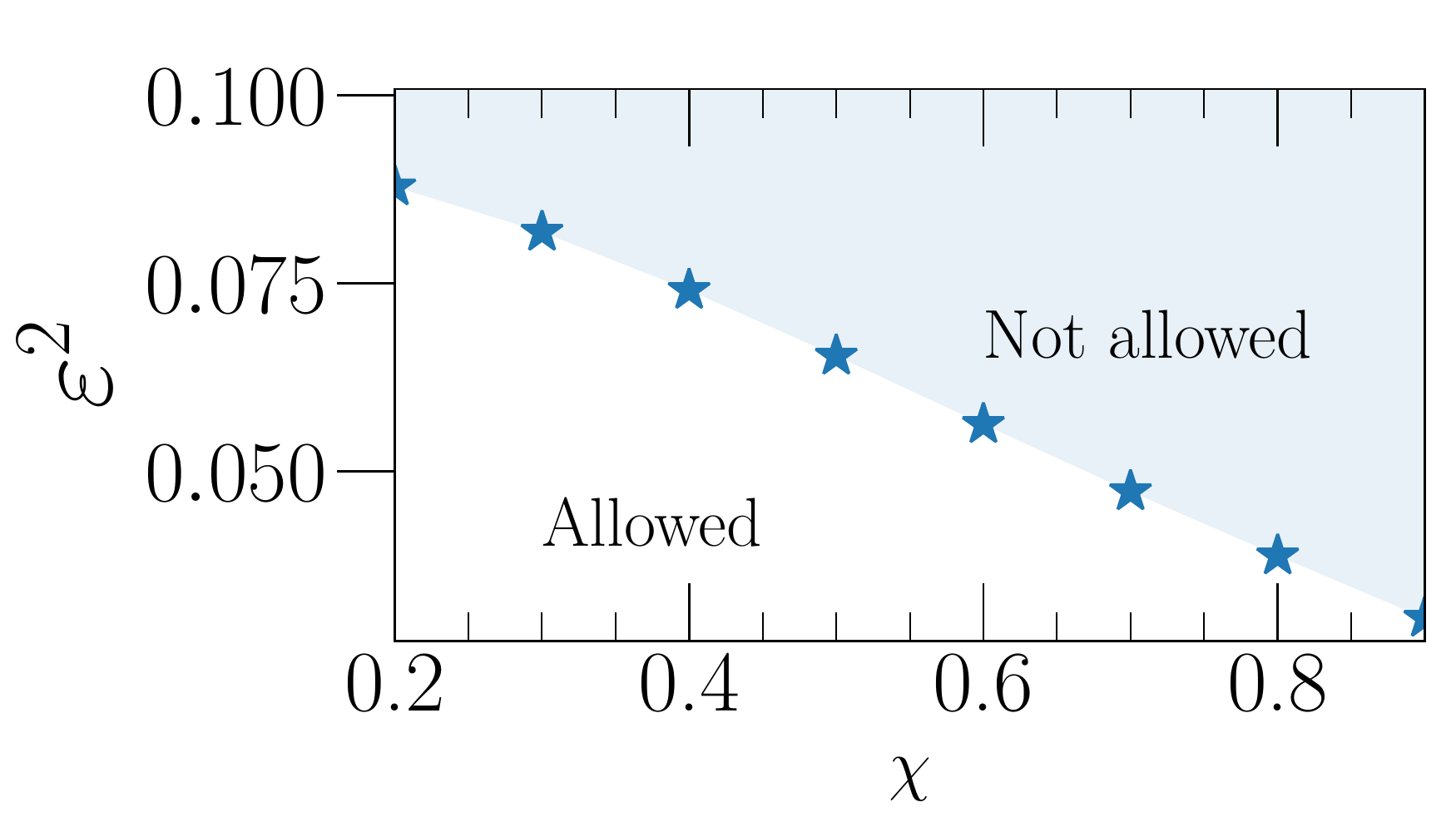}
  \caption{
Evaluation of the regime of validity as given by Eq.~\eqref{eq:validity}, for various values of spin. The top region is not allowed by perturbation theory, while the bottom region is allowed. The stars denote the values of $\chi$ at which we have evaluated Eq.~\eqref{eq:validity}. We can compare this to the regime of validity figure given in~\cite{Stein:2014xba}. 
  }
  \label{fig:validity}
\end{figure}

To second order, the perturbed metric takes the form
\begin{align}
\psi_{ab} \to \psi_{ab} + \varepsilon^2 \Delta \psi_{ab}
\end{align}
where $\varepsilon^2$ determines the amplitude of the metric perturbation. For the perturbative scheme to be valid, we require that $\| \psi_{ab} \| \gtrsim \|\varepsilon^2 \Delta \psi_{ab} \|$, where $\| \|$ denotes the L2 norm of the field. The values of $\varepsilon^2$ that satisfy this condition define the \textit{regime of validity}. We can measure this value of $\varepsilon^2$ by comparing the magnitudes of $\psi_{ab}$ and $\Delta \psi_{ab}$ as 
\begin{align}
\label{eq:validity}
\eps^2_\mathrm{max} = 0.1 \left(\| \frac{\psi_{ab}}{\Delta \psi_{ab}} \| \right)_\mathrm{min} \,.
\end{align}
Here the ratio is taken pointwise on the domain, we have chosen a constant $0.1$ for the comparison, and we find a global minimum (the minimum is close to the horizon, where the perturbation is the largest). We plot the results in Fig.~\ref{fig:validity}, where for lower spins larger values of $\varepsilon^2$ are allowed. Recall that $\varepsilon$ counts powers of $\ell^2/GM$, and thus we can map this regime of validity result to $\ell$ as well. 

\subsection{Black hole shadows}

One application of this initial data framework is to study modifications to the black hole shadow. Observing black hole shadows explores an entirely new scale of gravitational curvature and thus can test GR in a wholly new way~\cite{Baker:2014zba}. Since looking at the shadow effectively involves observing the behavior of test particles (photons) moving on geodesics in the spacetime, observing the shadows of stationary black holes serves as a \textit{metric test} of GR. 

\subsubsection{EHT capabilities and previous work}
\label{sec:EHT}

Let us first review the capabilities of the Event Horizon Telescope (EHT) for detecting black hole shadows. The EHT is a very long baseline interferometry array of radio telescopes around the world that aims to generate images of the black hole at the center of the Milky Way galaxy, Sgr A*, as well as that of the M87 galaxy, with horizon-scale resolution. Electromagnetic images show not the actual horizon, but the region external to the light ring at $3GM/c^2$, which serves as a probe of the black hole shadow~\cite{Falcke:2018uqa}. Resolving Sgr A* requires an angular resolution of $\mathcal{O}(10)$ microarcseconds ($\mu\mathrm{as}$)~\cite{Psaltis:2018xkc}. Once complete, the array should have resolutions of up 23 $\mu\mathrm{as}$ at 230 GHz and 15 $\mu\mathrm{as}$ at 345 GHz~\cite{Ricarte:2014nca}. The size of Sgr A*'s visible event horizon is predicted to be $\sim 50\, \mu\mathrm{as}$~\cite{Falcke:2013ola}, with the photon ring contributing to $1$ -- $10 \%$ of the total flux~\cite{2013MNRAS.432.2252D}.

Actually predicting what black hole images will look like for Sgr A* and M87, however, requires simulating the matter around the black hole using GRMHD simulations (cf.~\cite{Psaltis:2018xkc} for a review). However, as the shadow only depends on the black hole spacetime, the shadow is not affected by the presence of matter~\cite{Psaltis:2014mca}. Nevertheless, observing the shadow free from  obscuration due to the accretion onto the black hole (and gravitational lensing thereof) is a technical challenge. Additionally, interstellar scattering affects the resolution of the image~\cite{Psaltis:2018xkc}. In this study, we only consider null rays and the scalar field around a black hole otherwise in vacuum when probing the shadow, and thus do not include the matter effects.

How well can the edge of the shadow be detected? Psaltis et al.~\cite{Psaltis:2014mca} took advantage of the fact that the black hole shadow produces some of the steepest gradients in an image, and applied various edge-finding algorithms to locate the shadow. In practice, thus, it is possible to extract to an extent an edge corresponding roughly to the black hole shadow to within $\sim 9\%$, assuming a given scattering kernel. 

How well can current algorithms measure the properties of the black hole shadow of Sgr A*? Fig. 13 of Psaltis et al.~\cite{Psaltis:2018xkc} shows a combined posterior distribution for the black hole quadrupole moment $q$ and the black hole spin $a$ for a hypothetical observation of Sgr A*. If the black hole is Kerr, then there should be a unique point in this space for each mass and spin on the curve $q = -a^2$. EHT observations give a wide curve in the $q$-$a$ space, while constraints from spin measurements from stars and pulsars around Sgr A* provide tighter constraints. Nevertheless, the spin in this posterior can only be predicted to an accuracy of $\sigma_a \sim 0.1$.

Previous studies have calculated (without considering matter effects) black hole shadows in alternative theories of gravity (see~\cite{Falcke:2013ola} and~\cite{Psaltis:2018xkc} for a review). Additionally, Ref.~\cite{Johannsen:2016uoh} reviews the detectability of effective \textit{deviation parameters} from otherwise GR predictions. 

\subsubsection{Computing the shadow}

We now compute the second-order deviation to the black hole shadow in order-reduced dynamical Chern-Simons gravity. Recall that we have solved for a metric perturbation $\Delta \psi_{ab}$ around an isolated black hole of a given spin. We can then add it to the background metric $\psi_{ab}$ via a coupling parameter $\eps^2$ that lies in the regime of validity outlined in Sec.~\ref{sec:ROV}. The overall metric is thus
\begin{align}
\psi_{ab}^\mathrm{pert} \equiv \psi_{ab} + \eps^2 \Delta \psi_{ab}\,.
\end{align}
We compute the dCS black hole shadow in this metric, which will be correct to second order. Note that since we have solved for stationary data, we only need to evolve geodesics on one time slice to trace the shadow, as all of the slices will be the same. Note also that since the shadow is a physical observable, we do not need to worry about gauge effects. 

To probe the shadow, we use the geodesic integration methods (and corresponding code) outlined in~\cite{0264-9381-32-6-065002} and~\cite{PhysRevD.94.064008}. We refer the reader to those papers for a technical discussion. Schematically, we start geodesics from a camera some $C = \mathcal{O}(10)\,M$ away from the black hole, and integrate them backwards in time. The geodesics that make it to past null infinity (which we approximate as a distance of $2C$ from the black hole in order to avoid integrating geodesics to infinity) are labeled as not in the shadow, while the geodesics that converge onto the horizon determine the edge of the shadow. The code has built-in refinement, and with increasing resolution more geodesics are added along the shadow edge. 

\subsubsection{Analyzing the shadow}
\label{sec:shadowAnal}

We now present a novel way to analyze the black hole shadow as computed from evolving null geodesics. Note that there exist previously-proposed methods of analyzing the shadow~\cite{Abdujabbarov:2015xqa}. Given the shadow edge in the $x$-$y$ plane of the camera (also known as the \textit{image plane}), parameterized as two functions $x(\theta)$ and $y(\theta)$ where $\theta$ is the angle about some chosen center, we can Fourier decompose  the shadow edge as
\begin{align}
x(\theta) &= a_0 + \sum_{n = 1}^N a_n \cos(n \theta)\,, \\
y(\theta) &= b_0 + \sum_{n = 1}^N b_n \sin(n \theta)\,,
\end{align}
up to some number $N$ of fitting coefficients. We define a measure of the power in each Fourier mode as
\begin{align}
\label{eq:fdef}
f_n \equiv \sqrt{a_n^2 + b_n^2}\,.
\end{align}

In this procedure, one must take precautions in defining the axes and the origin for $\theta$. Suppose we have an image of a black hole shadow. For simplicity, assume that the spin axis has no component normal to the plane of the camera, but has some arbitrary orientation in that plane. Given such an image, we can find a line about which the image has a reflection symmetry. Let this be the $x$-axis (in the case of $\chi = 0$, we can take any axis).

\begin{figure}
  \centering
   \includegraphics[width=0.8\textwidth]{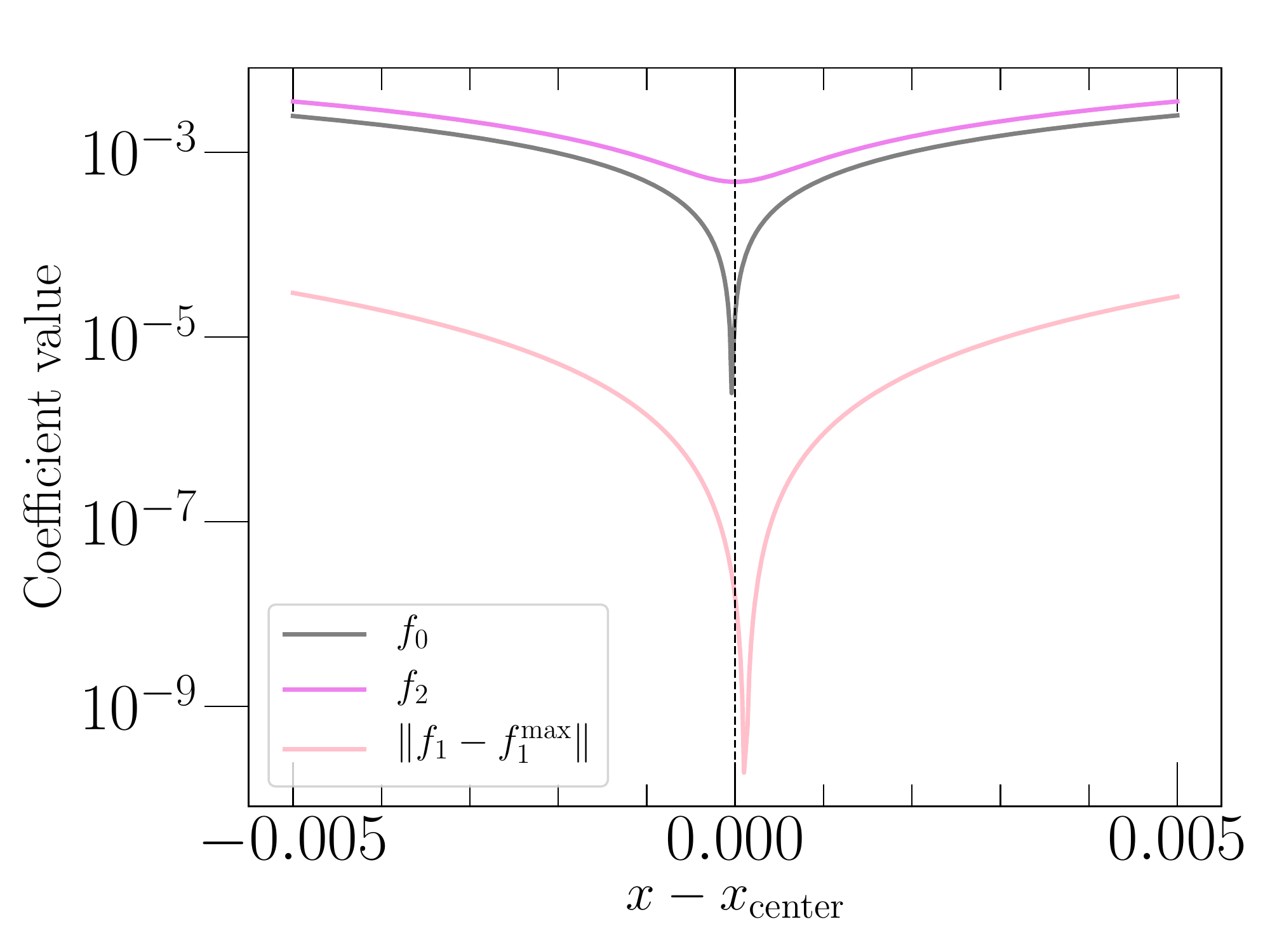}
  \caption{Results of our procedure for designating the center (and hence the origin for the angle $\theta$) of a black hole shadow for $\chi = 0.9$ and $\varepsilon^2=0$. We find the central value of $x$ by minimizing the recovered $n = 2$ multipole for each trial value. Here, we plot in the dashed black line the optimal value of $x$. We see that when $x$ is chosen to minimize $n = 2$, it also minimizes the artificial $n = 0$ multipole. Note that the minimum value of the $n = 2$ multipole is finite, as the shadow shape is non-spherical. Additionally, we plot the difference between the $n = 1$ multipole and its maximum value, finding that it attains the maximum near but not at the optimum center value as the shape is not exactly spherical. 
  }
  \label{fig:CenterFinding}
\end{figure}

Next, we need to define an origin $\{x_0, y_0\}$ in the $x$-$y$ plane from which to measure the angle $\theta$. For $y_0$, we can simply choose $y_0 = 0$ since we have defined $y = 0$ to be the axis of reflection symmetry.  For $x_0$, however, we need to be more careful. In the $\chi = 0$ case, for example, one can choose an $x_0$ such that the decomposition has an artificially non-zero $n = 2$ multipole. Thus, we choose $x_0$ to be the point at which the value of $f_2$ is minimized. We show the result of this procedure in Fig.~\ref{fig:CenterFinding}. 

\begin{figure}
  \centering
   \includegraphics[width=0.6\textwidth]{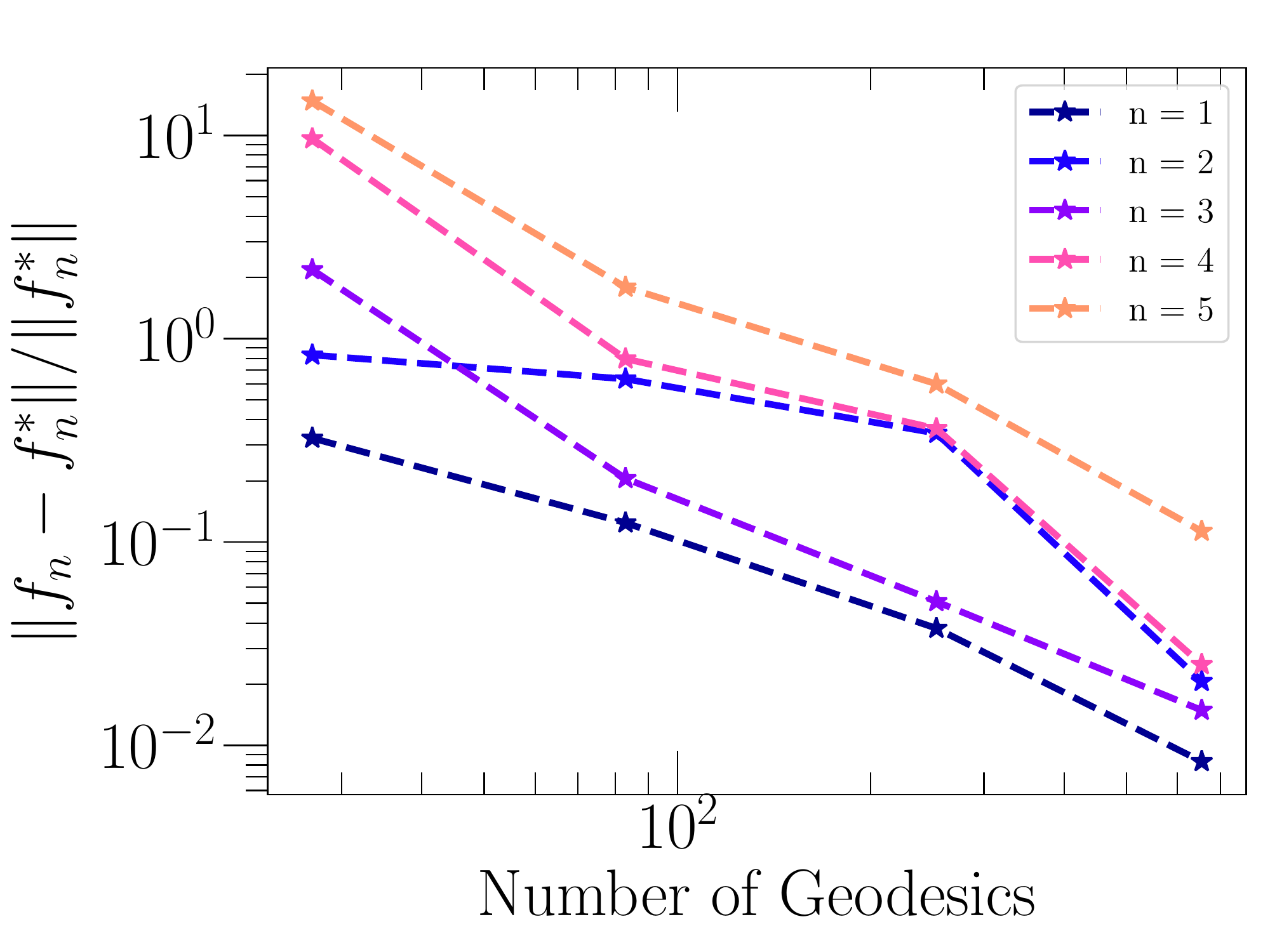}
  \caption{Convergence of the shadow multipoles with resolution for a spin $\chi = 0.9$ black hole for $\varepsilon^2=0$. For each multipole ($n = 1$ to $n = 5$), we plot the normalized difference of the value of the multipole from the highest resolution value (denoted as $f_n^*$), as a function of resolution. Here, the $x$-axis corresponds to the number of geodesics that converge onto the horizon when integrated backwards in time, and hence are used to image the black hole shadow. As we increase this resolution, the normalized differences from the highest resolution value decrease. We see that the higher multipoles, which take more geodesics to resolve, converge more slowly than the lower multipoles.  }
  \label{fig:Spin0p9CoeffConvergence}
\end{figure}

We also check that the values of the coefficients given in the decomposition~\eqref{eq:fdef} converge with resolution.  We show a quantitative convergence analysis in Fig.~\ref{fig:Spin0p9CoeffConvergence}. We check convergence for each shadow we analyze, for a given $\chi$ and $\varepsilon^2$.

The $n = 0$ multipole refers to the coordinate location of the shadow center in the plane of the camera, which is not gauge-invariant and hence not meaningful. The $n = 1$ multipole corresponds to the `size' of the shadow, and is proportional to both the mass of the black hole and the distance to the camera. Thus, the value of the $n = 1$ multipole is not meaningful as there is a mass-distance degeneracy. However, dividing all of the $n > 1$ multipoles $f_n$ by $f_1$ gives normalized values that are independent of the mass and distance, and in the $\varepsilon^2 = 0$ case, only dependent on the dimensionless spin. We have verified this numerically by changing the mass of the black hole, and checking that the normalized $n > 1$ coefficients remain the same. We thus focus out attention on the $n > 1$ multipoles normalized by $f_1$, which have physical meaning. 

Now, in the presence of a nonzero $\eps$, we still apply this same procedure (orienting on the axis of reflection symmetry, finding the center by minimizing $f_2$, then dividing through by $f_1$). Note that in this case, we expect the higher multipoles to have a different dependence on $\chi$ and now $\eps$. We will need to observe at least two multipoles to perform a consistency check with the $\eps = 0$ case, or to estimate $\eps$ and $\chi$ if we find $\eps\neq 0$.

\subsubsection{Results}

\begin{figure}
\centering
 \includegraphics[width=\textwidth]{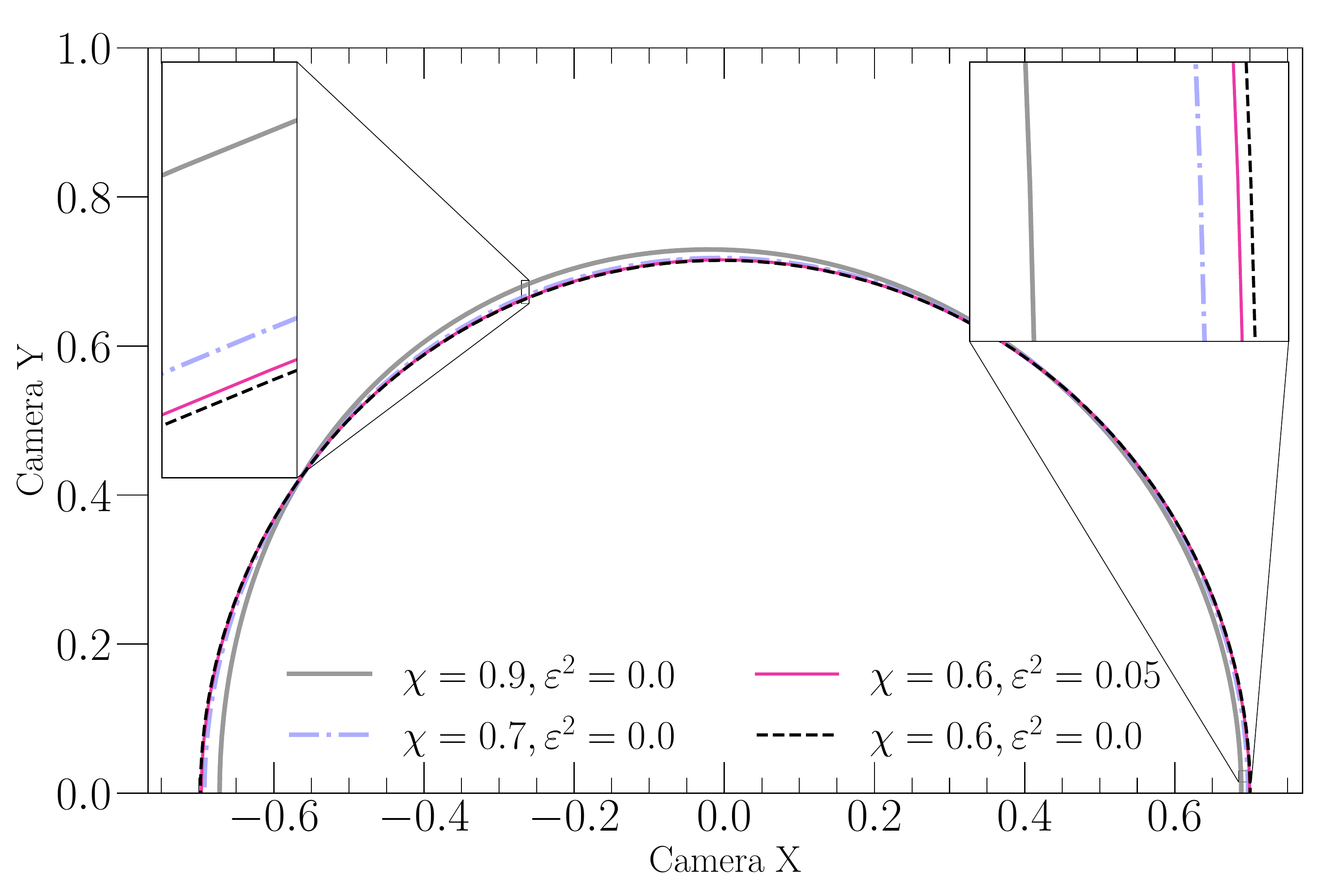}
  \caption{Visualization of black hole shadows. The $x$- and $y$-axes correspond to camera coordinates for a camera resolving the black hole, and thus are not physically meaningful. The shape of each shadow has been normalized by its overall `size' as given by the $n = 1$ multipole. Likewise, each shadow has been centered according the procedure described in this paper. We plot the shadow for spin of $\chi = 0.6$, with dCS perturbation parameters $\eps^2 = 0$ and $\eps^2 = 0.05$, the maximum allowed within the regime of validity. Zooming in, we see a difference in the two shadows. However, increasing the spin to $\chi = 0.7$ without a dCS perturbation (and even $\chi = 0.9$) has a stronger effect on the shape of the shadow. We have checked that increasing the resolution of the shadow by integrating more geodesics has a smaller effect than aforementioned the physical effects.}
  \label{fig:ShadowShape}
\end{figure}

Let us now analyze the black hole shadow using the procedure outlined in this section for various dimensionless spins $\chi$ of the background black hole and perturbation parameters $\varepsilon^2$. In accordance with the feasibility study shown in Fig.~13 of~\cite{Psaltis:2018xkc}, we concentrate our attention on spins of $\chi = 0.6$. In Fig.~\ref{fig:ShadowShape}, we  plot the black hole shadow for $\chi = 0.6$ for $\eps^2 = 0$ (i.e., the shadow as predicted by GR) and $\eps^2 = 0.05$, the maximal value allowed by the regime of validity. Additionally, we plot the GR shadows for $\chi = 0.7$ and $\chi = 0.9$ black holes. We see that shifting the spin away from $0.6$ has a greater effect than adding a dCS perturbation. Given the $\sigma_a \sim 0.1$ spread in the recovered spin for the trial EHT measurement in Ref.~\cite{Psaltis:2018xkc}, it is informative to compare the effect of increasing $\chi$ by $0.1$ versus increasing $\eps^2$ to its maximum valid value at a given $\chi$. We see that the effect increasing $\eps^2$ on the visual shape of the shadow is less than the effect from increasing $\chi$. 


\begin{figure}
  \centering
  \subfloat{\includegraphics[width=0.45\textwidth]{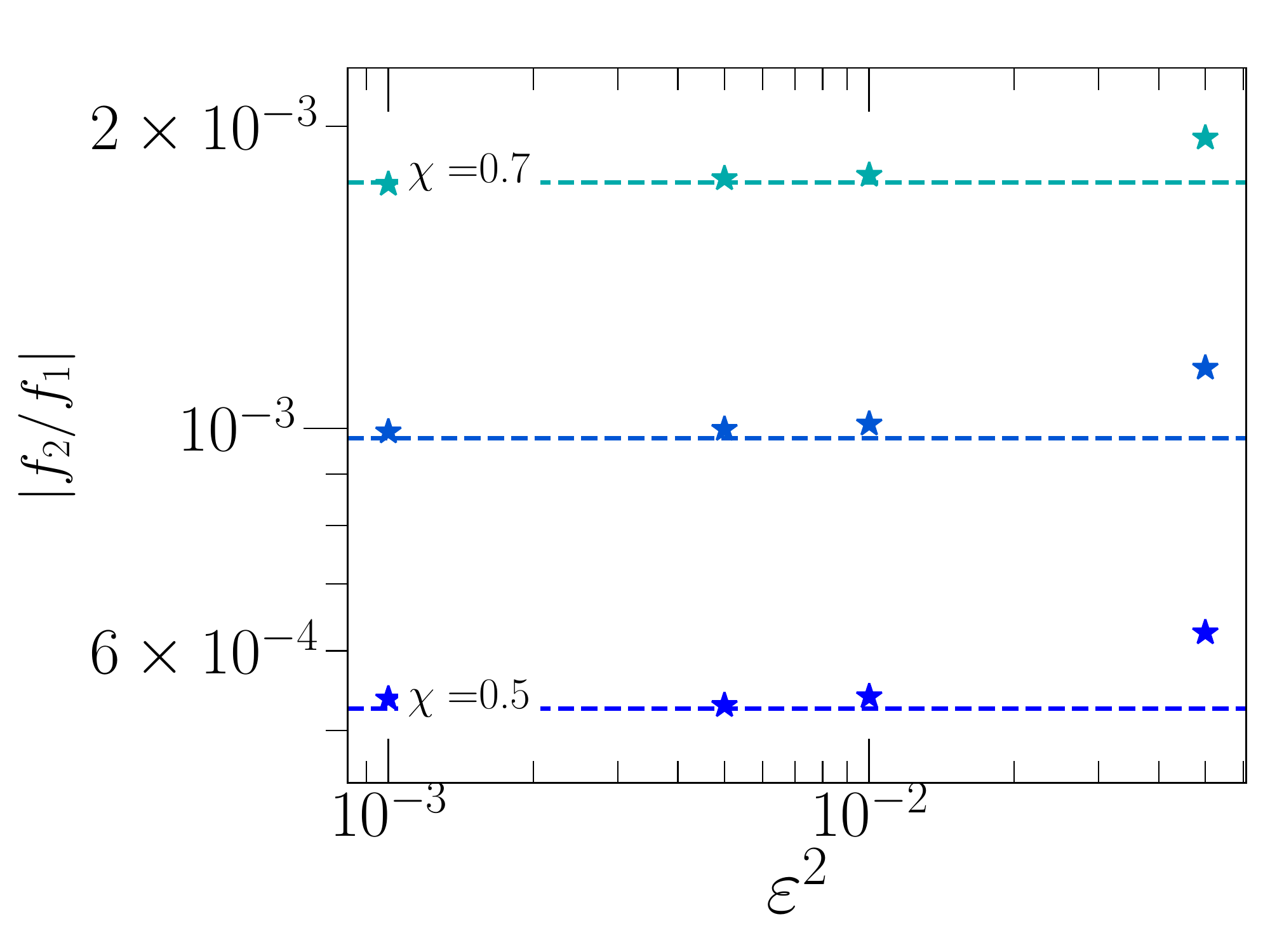} 
  \label{fig:SpinCoeffs2}}
  \subfloat{\includegraphics[width=0.45\textwidth]{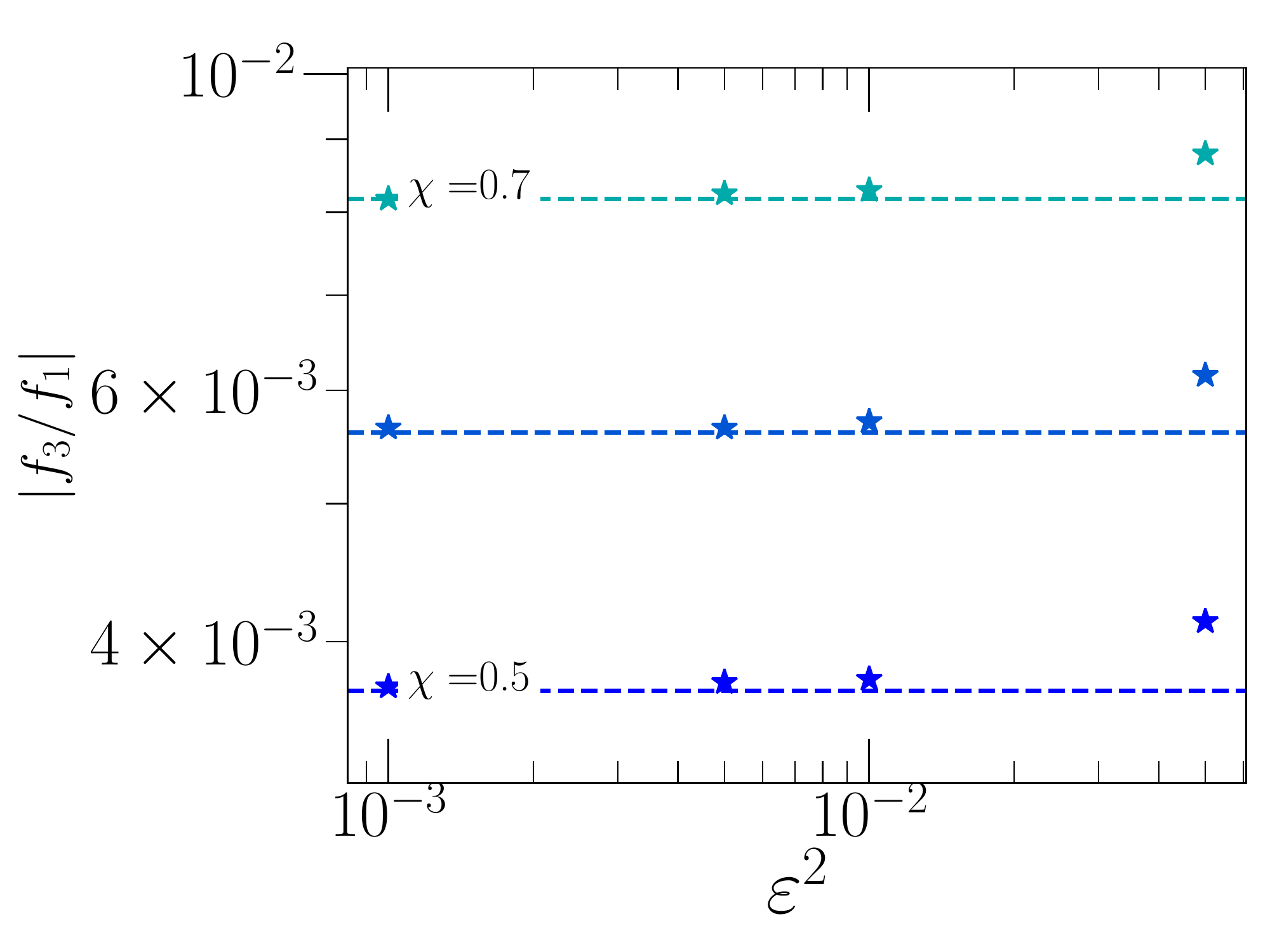} 
  \label{fig:SpinCoeffs3}} 
  \caption{
  Values of the $f_2$ and $f_3$ coefficients of the black hole shadow, as calculated using the methods outlined in Sec.~\ref{sec:shadowAnal}. Each coefficient is normalized by $f_1$, which corresponds to the size of the shadow. Each dashed line shows the $\eps^2 = 0$ value of the coefficient, corresponding to an unperturbed GR black hole, for spins $\chi = 0.5$, $\chi = 0.6$, and $\chi = 0.7$ (as labeled on the plot). Since the shadow in GR becomes less spherical with increasing spin, it is consistent that the $f_2$ and $f_3$ coefficients, which correspond to non-spherical multipoles, increase with spin. For each spin, we also plot the values of the multipoles when we introduce a dCS perturbation of the form $\psi_{ab} + \eps^2 \Delta \psi_{ab}$. As we increase $\eps^2$ (up to a value given by the regime of validity of perturbation theory), we see that these coefficients increase as well, in a power-law fashion. We have checked that increasing the resolution of the shadow by integrating more geodesics leads to convergent results for the multipoles, and does not affect the results on the scale presented here.} 
\end{figure}

We can quantitatively analyze the shape of the shadow by considering the values of $f_2/f_1$ and $f_3/f_1$, the two dominant normalized multipoles. Considering again spins around $\chi = 0.6$, we plot the values of these multipoles with increasing $\eps^2$ in Figs.~\ref{fig:SpinCoeffs2} and~~\ref{fig:SpinCoeffs3}. We see that, for a given spin, as we increase $\eps^2$, the values of $f_2/f_1$ and $f_3/f_1$ linearly deviate away from the $\eps^2 = 0$, GR prediction. Since the shadow, with the mass normalized away, is dependent only on $\chi$ and $\eps^2$ in dCS, we can map
\begin{align}
\label{eq:chiMap}
\{ \chi, \eps^2 \} \to \{ f_2/f_1, f_3/f_1 \}\,,
\end{align}
for each choice of $\chi$ and $\eps^2$. 

While the mapping shown in Eq.~\eqref{eq:chiMap} is unique for each $\{ \chi, \eps^2 \}$ pair, it may not be invertible. In other words, degeneracies may exist such that a given pair $\{ f_2/f_1, f_3/f_1\}$ can be generated by more than one combination of $\{\chi, \eps^2\}$. In particular, this degeneracy can spoil a GR null hypothesis test using the shadow. Suppose there exists a spin $\chi_a$ and $\eps_a^2 \neq 0$ combination such that the corresponding $f_2/f_1$ and $f_3/f_1$ values are equal to those of a $\chi_b$ and $\eps_b^2 = 0$ shadow. Then, we would not be able to distinguish a black hole with a dCS perturbation from a Kerr black hole with a different spin. 

\begin{figure*}
  \centering
   \includegraphics[width=\textwidth]{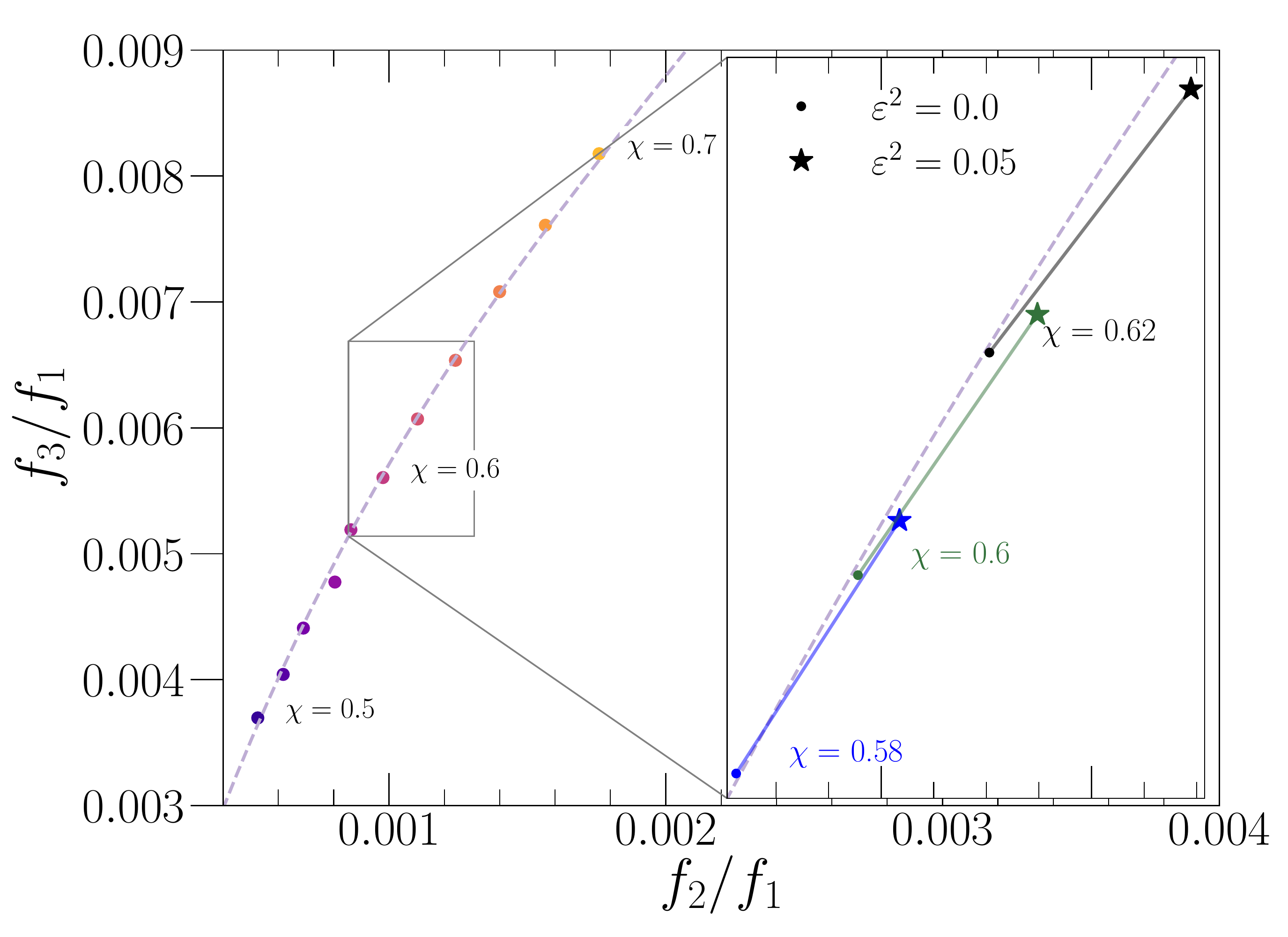}
  \caption{
 Normalized black hole shadow coefficients of the $n = 2$ ($x$-axis) and $n = 3$ ($y$-axis) multipoles. These correspond to the dominant non-spherical multipoles. The figure explores degeneracies in the $\chi$-$\eps$ space. In the \textbf{(left)} panel, we plot the coefficient values for $\eps^2 = 0$ for a variety of spins from $\chi = 0.5$ to $\chi = 0.7$. We additionally plot a curve (dashed line) that we have fit for all of the $\eps^2 = 0$ coefficient values over a broader range of spins ($\chi = 0.1$ to $\chi = 0.9$). This line is the Kerr curve in the $f_2$-$f_3$ space. In the \textbf{(right)} panel, we introduce dCS perturbations. We plot again the Kerr curve, and consider coefficient values for spins of $0.58$, $0.6$, and $0.62$. We see that when we introduce a dCS perturbation of strength $\eps^2 = 0.05$, the values of the coefficients deviate from the Kerr curve. The fact that the perturbed values do not lie on the Kerr curve gives us a handle on the amount of degeneracy in the $\chi$-$\eps$ space. We have checked that these effects are convergent with increasing the resolution of the shadow by integrating more geodesics.}
  \label{fig:Degeneracies}
\end{figure*}

We explore this potential degeneracy in Fig.~\ref{fig:Degeneracies}. Using the $\eps^2 = 0$ values of $f_2 / f_1$ and $f_3/f_1$ for various spins, we trace out a curve in this multipolar parameter space. This curve is solely parametrized by spin $\chi$, and any deviation away from this curve corresponds to some additional, non-Kerr effects. We call this the `Kerr' curve. Then, considering $\chi = 0.6$ and neighboring spins, we consider the effect of adding an $\eps^2 = 0.05$ dCS perturbation. We see that in the presence of $\eps^2 \neq 0$, the multipolar values deviate away from the Kerr curve. In other words, we do not have a $\chi$-$\eps^2$ degeneracy. This in turn makes a GR null-hypothesis test possible using dCS shadows. On the other hand, we can also see from the figure that it may be difficult to distinguish various $\{\chi, \eps^2 \neq 0\}$ pairs. However, since $\eps^2$ is a universal parameter, observing more and more black hole shadows in practice should statistically narrow the value. 

Let us now consider these results in the context of the EHT capabilities outlined in Sec.~\ref{sec:EHT}. We claim, given our investigation of the shape of the shadow, that precisely quantifying $\chi$ and $\eps^2$ for Sgr A*, for example, may be infeasible with the current EHT resolution. Given that observations can yield a spread of as much as $0.2$ in the spin, and given that we have seen that dCS effects for the maximum allowed values of $\eps^2$ are smaller than a $0.1$ increase in the spin, it will be difficult to observe such a deviation with the EHT. However, increasing the resolution of EHT shadow edge observations will allow us to perhaps probe these small effects, in part to perform an analysis to check for $\eps^2 = 0$ consistency, or at  least bound large values of $\eps^2$. 

Suppose that an external measurement of the mass of Sgr A* was available. Because the size of the shadow on the camera depends not only on mass but also on distance, we would need to have a measurement of the distance of Sgr A* as well. In this case, we would not need to normalize all of $f_{> 1}$ coefficients by $f_1$, since the mass would be known. However, the $f_1$ multipole is independent of spin, and thus a deviation of the $f_1$ multipole predicted from an independent measurement of the mass and distance of Sgr A* could point to a non-GR signature. Such an analysis was performed, for example in~\cite{Psaltis:2014mca}.

On the other hand, suppose there were an independent measurement of the spin of Sgr A* from pulsars~\cite{Psaltis:2010ca}, with tighter constraints than the example shown in Ref.~\cite{Psaltis:2018xkc}. If $\chi$ was known precisely from pulsar measurements, then we would simply use the value of the dominant multipole $f_3$ to observe deviations from the predicted value in the case of $\eps^2 = 0$. Fig.~\ref{fig:SpinCoeffs3} shows the value of $f_3$ away from its predicted GR value for a spin of $\chi = 0.6$, for example. Knowing precisely the value of $\chi$ would thus allow us to constrain the value of $\eps^2$ in the $\eps^2$ - $f_3$ space. However, we must be careful in noting that this would only serve as a null-hypothesis test of GR, as inferring $\chi$ from pulsar measurements (presently) assumes that GR is the underlying model. 

\section{Conclusion}
\label{sec:Conclusion} 

In this paper, we have presented a method for numerically generating metric perturbation initial data (Sec.~\ref{sec:Solving}), applied it to dynamical Chern-Simons gravity (Sec.~\ref{sec:dCSData}), and investigated black hole shadows in the presence of dCS metric perturbations (Sec.~\ref{sec:dCSPhysics}). 

The metric perturbation initial data computation is fully general, meaning that given some metric perturbation source, background spacetime, and boundary conditions (as well as specifying a choice of the free data), we can produce constraint-convergent first-order metric perturbation results. In particular, we can easily extend the dCS initial data results for a single black hole presented in this paper to the binary case. We can also, for example, apply this initial data formalism to explore linear versus non-linear metric perturbations in a standard Kerr spacetime, as our metric perturbation data is constraint-satisfying to first order (for example, to compare to the metric perturbation data used in~\cite{East:2013mfa} and~\cite{Bhagwat:2017tkm}).

Future work in this program involves evolving dCS initial metric perturbations. This is done following the order-reduction scheme (cf. Sec.\ref{sec:OrderReduction} and~\cite{Okounkova:2017yby}), which guarantees well-posedness, as each order in the scheme has the same principal part as the general relativity background. One possibility is to evolve a single spinning black hole to see if it is stable. A second is for the binary black hole case. There, we can evolve the metric perturbation sourced by the dCS scalar field and generate perturbed gravitational waveforms, thus performing the next step of the program outlined in~\cite{Okounkova:2017yby}. 

\section*{Acknowledgements}

We would like to thank Leo Stein for many useful discussions and providing us with the code used to generate the initial data in~\cite{Stein:2014xba}. We would also like to thank Francois Hebert for helping us use the SXS lensing code. We thank Harald Pfeiffer and Leo Stein for careful reading of this manuscript. This work was supported in part by the Sherman Fairchild Foundation, and NSF grants PHY-1708212 and PHY-1708213 at Caltech and PHY-1606654 at Cornell. Computations were performed using the Spectral Einstein Code~\cite{SpECwebsite}, including the Simulating Extreme Spacetimes collaboration lensing code~\cite{0264-9381-32-6-065002}. All computations were performed on the Wheeler cluster at Caltech, which is supported by the Sherman Fairchild Foundation and by Caltech.

\appendix
\section{Perturbed extended conformal thin sandwich quantities}
\label{sec:PerturbedQuantitiesAppendix}

In this appendix, we derive the first-order perturbations to all of the extended conformal thin sandwich quantities, which enter into Eqs.~\eqref{eq:DeltaPsiEq},~\eqref{eq:DeltaShiftEq}, and~\eqref{eq:DeltaCEq}.

First, the perturbation to the inverse of the conformal spatial metric is 
\begin{align}
\Delta \bar{g}^{ij} = -\bar{g}^{ik} \bar{g}^{jm} \Delta \bar{g}_{km} \,.
\end{align}
We can use this to obtain the useful identities
\begin{align}
\Delta V_i &= \Delta \bar{g}_{ij} V^{j} + \bar{g}_{ij} \Delta V^{j}  \,,  \\
\Delta F^{kl} &= \Delta \bar{g}_{ki} \bar{g}_{lj} F^{ij} + \bar{g}_{ki} \Delta \bar{g}_{lj} F^{ij}  +\bar{g}_{ki} \bar{g}_{lj}  \Delta F^{ij} \,, \\
\label{eq:PertTrace}
\Delta F &= \Delta \bar{g}^{ij} F_{ij} + \bar{g}^{ij} \Delta F_{ij} \,,
\end{align}
for vectors $V^i$ with perturbation $\Delta V^i$ and tensor $F_{ij}$ with trace $F$ and perturbation $\Delta F_{ij}$. 

The covariant derivative operator $\bar{D}$ will also have a perturbation. We perturb the Christoffel symbols corresponding to $\bar{g}_{ij}$ to obtain
\begin{align}
\Delta \bar{\Gamma}^i_{jk} &= \frac{1}{2}\Delta \bar{g}^{il}(\pd_k \bar{g}_{lj} + \pd_j \bar{g}_{lk} - \pd_l \bar{g}_{jk}) + \frac{1}{2}\bar{g}^{il}(\pd_k \Delta \bar{g}_{lj} + \pd_j \Delta \bar{g}_{lk} - \pd_l \Delta \bar{g}_{jk})\,.
\end{align}
This in turn gives the useful perturbed derivative identities
\begin{align}
\Delta (\bar{D)}_i S& = 0 \,, \\
\Delta (\bar{D})^i S &= \Delta \bar{g}^{ij} \bar{D}_j S \,, \\
\Delta (\bar{D}^2) S &= \Delta \bar{g}^{ij} \pd_i \pd_j S -\Delta \bar{g}^{ij} \bar{\Gamma}^l_{ij} \pd_l S  - \bar{g}^{ij} \Delta \bar{\Gamma}^l_{ij} \pd_l S\\
\Delta (\bar{D})_i V^j &= \Delta \bar{\Gamma}^i_{jk} V^k \,, \\ 
\Delta (\bar{D})_i V_j &= \Delta \bar{\Gamma}^k_{ij} V_k \,, \\
\Delta (\bar{D})^i V^j & = \Delta \bar{g}^{ik}  \bar{D}_k V^j + \bar{g}^{ik}  \Delta \bar{\Gamma}^j_{kl} V^l \,, \\
\Delta (\bar{D})_k F_{ij} &= -\Delta \bar{\Gamma}^m_{ki} F_{mj} - \Delta \bar{\Gamma}^m_{kj} F_{im} \,, \\
\Delta (\bar{D})_k F^{ij} &= \Delta \bar{\Gamma}^i_{km} F^{mj} +\Delta \bar{\Gamma}^j_{km} F^{im} \,,
\end{align}
for any scalars $S$ with perturbation $\Delta S$, vectors $V^i$, with perturbation $\Delta V^i$, and tensor $F_{ij}$, with perturbation $\Delta F_{ij}$. The parenthesis in expressions such as $\Delta (\bar{D}^2) S$ refer to the perturbation on just the derivative operator. 

Then we can compute the perturbation to the spatial Ricci tensor as
\begin{align}
\Delta \bar{R}_{ij} &= \pd_m \Delta \bar{\Gamma}^m_{ij}  - \frac{1}{2}(\pd_i \Delta \bar{\Gamma}^m_{mj} + \pd_j \Delta  \bar{\Gamma}^m_{mi}) \\
\nn \quad& + \Delta \bar{ \Gamma}^m_{ij} \bar{\Gamma}^n_{nm} - \Delta \bar{\Gamma}^m_{in} \bar{\Gamma}^n_{mj} + \bar{ \Gamma}^m_{ij} \Delta \bar{\Gamma}^n_{nm} - \bar{\Gamma}^m_{in} \Delta \bar{\Gamma}^n_{mj} \,.
\end{align}
and $\Delta \bar{R}$ can then be computed using Eq.~\eqref{eq:PertTrace}. 

Meanwhile, the perturbation to $\bar{L} \beta^{ij}$, defined in Eq.~\eqref{eq:LShift}, is 

\begin{align} 
\Delta (\bar{L} \beta)^{ij} &= \Delta( \bar{D})^i \beta^j  + \bar{D}^i \Delta \beta^j  +  \Delta( \bar{D})^j \beta^i +  \bar{D}^j \Delta \beta^i  \\
\nn & \quad - \frac{2}{3} \Delta \bar{g}^{ij} \bar{D}_k \beta^k  - \frac{2}{3} \bar{g}^{ij} (\Delta (\bar{D})_k \beta^k + \bar{D}_k \Delta \beta)^k\,.
\end{align}
For simplicity, we can group the terms with the background derivative operators operating on $\Delta \beta^i$, defining
\begin{align}
\label{eq:DeltaLShiftSeparation}
\Delta (\bar{L} \beta)^{ij} &= (\bar{L} \Delta \beta)^{ij}  + (\Delta{\bar{L}} \beta)^{ij} \,,
\end{align}
where
\begin{align}
(\bar{L} \Delta \beta)^{ij} \equiv \bar{D}^i \Delta \beta^j + \bar{D}^j \Delta \beta^i - \frac{2}{3} \bar{g}^{ij} \bar{D}_k \Delta \beta^k \,,
\end{align}
and
\begin{align}
(\Delta(\bar{L}) \beta)^{ij} &\equiv  \Delta (\bar{D})^i \beta^j +  \Delta (\bar{D})^j \beta^i    - \frac{2}{3} \Delta \bar{g}^{ij} \bar{D}_k \beta^k  - \frac{2}{3} \bar{g}^{ij} \Delta (\bar{D})_k \beta^k \,.
\end{align}

Finally, the perturbation to $\bar{A}^{ij}$, defined in Eq.~\eqref{eq:Aij}, is 
\begin{align}
\Delta \bar{A}^{ij} &=  7 \frac{\psi^6 \Delta \psi}{2\alpha \psi} ((\bar{L}\beta)^{ij} - \bar{u}^{ij})-\frac{\psi^7}{2(\alpha \psi)^2} \Delta C ((\bar{L}\beta)^{ij} - \bar{u}^{ij}) \\
\nn & + \frac{\psi^7}{2\alpha \psi} (\Delta(\bar{L}\beta)^{ij} - \bar{\Delta u}^{ij}) \,.
\end{align}

The perturbations to the source terms given in Eqs.~\eqref{eq:rho}~\eqref{eq:Si}~\eqref{eq:Sij} and~\eqref{eq:S} are
\begin{align}
\label{eq:DeltaRho}
\Delta \rho &\equiv \Delta n_a n_b T^{ab} + n_a \Delta n_b T^{ab} + n_a n_b \Delta T^{ab} \,, \\
\label{eq:DeltaSi}
\Delta S^i &\equiv -\Delta g^{ij} n^a T_{aj}  -g^{ij} \Delta n^a T_{aj}  -g^{ij} n^a \Delta T_{aj} \,, \\
\label{eq:DeltaSij}
\Delta S_{ij} &\equiv \Delta g_{ia} g_{jb} T^{ab}  + g_{ia} \Delta g_{jb} T^{ab} + g_{ia} g_{jb} \Delta T^{ab} \,, \\
\label{eq:DeltaS}
\Delta S &\equiv \Delta g^{ij} S_{ij} + g^{ij} \Delta S_{ij} \,.
\end{align}
For a vacuum background ($T_{ab} = 0$), these terms simplify to give
\begin{align}
\Delta \rho &\equiv n_a n_b \psi^{ac} \psi^{bd} \Delta T_{cd} = n^a n^b \Delta T_{ab} \,, \\
\Delta S^i &\equiv -g^{ij} n^a \Delta T_{aj} \,, \\
\Delta S_{ij} &\equiv \Delta T_{ij} \,, \\
\Delta S &\equiv g^{ij} \Delta S_{ij} \,.
\end{align}
Note that all of the above terms use the background variables without applying a conformal transformation. 

\section{Reconstructing the perturbed spacetime metric}
\label{sec:ReconstructionAppendix}

In this appendix, we detail how to reconstruct the (non-conformal) spatial metric, $\Delta g_{ij}$, and its time derivative, $\pd_t \Delta g_{ij}$, from the perturbed extended conformal thin sandwich variables solved for in Sec.~\ref{sec:PerturbedFormalism}. This in turn allows us to construct the perturbation to the spacetime metric, $\Delta \psi_{ab}$, and its time derivative, $\pd_t \Delta \psi_{ab}$.

We obtain, perturbing Eq.~\eqref{eq:gConformal}

\begin{align}
\Delta g_{ij} = \psi^4 \Delta \bar{g}_{ij} + 4 \psi^3 \Delta \psi \bar{g}_{ij}\,,
\end{align}
and
\begin{align}
{\Delta g}^{ij} =\psi^{-4} \Delta \bar{g}_{ij} - 4 \psi^{-5} \Delta \psi \bar{g}^{ij}\,.
\end{align}

For $u_{ij}$, we perturb Eq.~\eqref{eq:uConformal} to give
\begin{align}
\Delta u_{ij} = \psi^4 \Delta \bar{u}_{ij} + 4 \psi^3 \Delta \psi \bar{u}_{ij}\,,
\end{align}
which is in turn related to $\pd_t \Delta g_{ij}$ through perturbing Eq.~\eqref{eq:dtgConformal} to give
\begin{align}
\label{eq:DeltaUReconstruction}
\Delta u_{ij} &=  \pd_t \Delta g_{ij} - \frac{2}{3} \Delta g_{ij} (-\alpha K + D_i \beta^i) \\
\nn & \quad - \frac{2}{3} g_{ij} (-\Delta \alpha K + \Delta (D)_i \beta^i - \alpha \Delta K + D_i \Delta \beta^i) \,.
\end{align}

Finally, the perturbed extrinsic curvature $\Delta K_{ij}$ can be reconstructed from $\Delta K$ and the solved variables 
following Eqs.~\eqref{eq:KDecomposed} and~\eqref{eq:AConformal} as 
\begin{align}
\Delta K_{ij} = \Delta A_{ij} + \frac{1}{3} (\Delta g_{ij} K + g_{ij} \Delta K) \,,
\end{align}
where 
\begin{align}
\Delta A_{ij} = \psi^{-2} \Delta \bar{A}_{ij}  - 2 \psi^{-3} \Delta \psi \bar{A}_{ij} \,.
\end{align}

In addition to $\Delta g_{ij}$ and $\pd_t \Delta g_{ij}$, some applications, such as computing the black hole shadow, require the perturbation to the full spacetime metric $\psi_{ab} \to \psi_{ab} + \Delta \psi_{ab} $ and its time derivative $\pd_t \Delta \psi_{ab}$. We thus construct the spacetime metric perturbation as 
\begin{align}
\label{eq:Deltapsiab}
\Delta \psi_{ab} =  \bigg(\begin{matrix}
-2\alpha \Delta \alpha +  \Delta \beta_m \beta^m + \beta_m \Delta \beta^m & \Delta \beta_i \\
\Delta \beta_j & \Delta g_{ij}
\end{matrix}\bigg)\,.
\end{align}

For the time derivative, given by applying the chain rule to the terms in Eq.~\eqref{eq:Deltapsiab}, we need to specify the time derivatives of $\beta^i$, $\alpha$, $\Delta \beta^i$, and $\Delta \alpha$. For the background case, we can freely specify $\pd_t \beta^i = 0$ and $\pd_t \alpha = 0$~\cite{baumgarteShapiroBook}. We can apply the same principle to the perturbed data, and freely set $\pd_t \Delta \alpha = 0$ and $\pd_t \Delta \beta^i = 0$. For a stationary background ($\pd_t \psi_{ab} = 0$, where $\pd_t$ is a linear combination to Killing vector fields), we obtain
\begin{align}
\pd_t &(\Delta \beta_m \beta^m + \beta_m \Delta \beta^m) \\
\nn &= \pd_t (\Delta g_{mi} \beta^i \beta^m + g_{mi} \Delta \beta^i \beta^m + g_{mi} \beta^i \Delta \beta^m) \\
\nn &= \pd_t \Delta g_{mi} \beta^i \beta^m \,,
\end{align}
and thus
\begin{align}
\pd_t \Delta \psi_{ab} =  \bigg(\begin{matrix}
 \pd_t \Delta g_{ij} \beta^i \beta^j & \pd_t \Delta g_{ij} \beta^j \\
\pd_t \Delta g_{ij} \beta^i &\pd_t \Delta g_{ij}
\end{matrix}\bigg)\,.
\end{align}

\section*{References}

\bibliography{dCS_paper.bib}

\end{document}